\documentclass[onecolumn]{article}
\usepackage[english]{babel}

\usepackage[utf8]{inputenc}
\usepackage[pdftex]{graphicx}
\usepackage{tabularx}
\usepackage{authblk}
\usepackage{dcolumn}
\usepackage{amsmath}
\usepackage{longtable}
\usepackage{tabularx}
\usepackage{setspace}
\usepackage{multirow}
\usepackage{mathtools}
\usepackage{tikz}
\usepackage{lscape}
\usepackage{amssymb}
\usepackage{url}
\usepackage{arydshln}
\usepackage{lscape}
\usepackage[misc]{ifsym}
\usepackage{apalike}
\usepackage{hyperref}

\newcommand{\beginsupplement}{%
        \setcounter{table}{0}
        \renewcommand{\thetable}{S\arabic{table}}%
        \setcounter{figure}{0}
        \renewcommand{\thefigure}{S\arabic{figure}}%
        \setcounter{equation}{0}
        \renewcommand{\theequation}{S\arabic{equation}}%
        \setcounter{section}{0}
        \renewcommand{\thesection}{S\arabic{section}}%
     }

\usepackage{csquotes}
\MakeOuterQuote{"}
\usepackage{lscape}

\usepackage{lineno}

\usepackage[resetlabels]{multibib}
\newcites{latex}{Supplementary references}

\usepackage[font=small]{caption}

\RequirePackage[twoside,letterpaper,includeheadfoot,layoutsize={8.125in,10.875in},layouthoffset=0.1875in,layoutvoffset=0.0625in,left=38.5pt,right=43pt,top=43pt,bottom=32pt,headheight=0pt,headsep=10pt,footskip=25pt]{geometry}
\title{Disadvantaged students increase their academic performance through collective intelligence exposure in emergency remote learning due to COVID 19.}

\author[1,2,3 \Letter]{\small Cristian Candia}

\author[2,4, \Letter]{Alejandra Maldonado-Trapp}

\author[2]{Karla Lobos}

\author[2]{Fernando Pe\~na}

\author[2,5]{Carola Bruna}

\affil[1]{Data Science Institute, Facultad de Ingeniería, Universidad del Desarrollo, Las Condes, 7610658, Chile.}
\affil[2]{Laboratorio de Investigación e Innovación Educativa, IDECLAB, Dirección de Docencia, Universidad de Concepción, Concepción, 4070383, Chile}

\affil[3]{Northwestern Institute on Complex Systems (NICO), Northwestern University, Evanston, IL 60208, United States}
\affil[4]{Departamento de Física, Facultad de Ciencias Físicas y Matemáticas, Universidad de Concepción, Concepción, Chile}
\affil[5]{Departamento de Bioquímica y Biología Molecular, Facultad de Ciencias Biológicas, Universidad de Concepción, Concepción, Chile}

\date{\today}   

\begin{document}
\maketitle

\begin{abstract}
During the COVID-19 crisis, educational institutions worldwide shifted from face-to-face instruction to emergency remote teaching (ERT) modalities. In this forced and sudden transition, teachers and students did not have the opportunity to acquire the knowledge or skills necessary for online learning modalities implemented through a learning management system (LMS). Therefore, undergraduate teachers tend to mainly use a LMS as an information repository and rarely promote virtual interactions among students, thus limiting the benefits of collective intelligence for students. We analyzed data on 7,528 undergraduate students and found that cooperative and consensus dynamics among university students in discussion forums positively affect their final GPA, with a steeper effect for students with low academic performance during high school. These results hold above and beyond socioeconomic and other LMS activity confounders. Furthermore, using natural language processing, we show that first-year students with low academic performance during high school are exposed to more content-intensive posts in discussion forums, leading to significantly higher university GPAs than their low-performance peers in high school. We expect these results to motivate higher education teachers worldwide to promote cooperative and consensus dynamics among students using tools such as forum discussions in their classes to reap the benefits of social learning and collective intelligence.
\end{abstract}

\section*{\label{sec:intro}Main}
We are constantly exposed to collective intelligence that, broadly speaking, aims at the enrichment of individuals \cite{levy1997collective}. Examples include cooperating and reaching consensus in team tasks, searching for past customers’ experiences, and accessing nonlocal and distributed knowledge on Google or Wikipedia. Integrating information spread among a myriad of people to make it available to the public upon a single query is a type of collective intelligence that aims to improve individual outcomes \cite{levy1997collective}. However, little is known about the benefits of this type of collective intelligence in educational experiences.

The recent COVID-19 crisis forced a transition from face-to-face classes to an emergency remote teaching (ERT) modality with the assistance of the digital technologies of a Learning Management System (LMS). In this pressing process, teachers and students did not have the opportunity to acquire the knowledge or skills to respond to the remote teaching modality. Therefore, teachers tend to mainly use a LMS as an information repository and rarely promote virtual interactions among students, thus limiting the benefits of collective intelligence for students. In addition, students had to adopt new ways to communicate and interact with each other for academic purposes, such as email, chat rooms, and academic forums. Thus, the emergency remote learning modality provides a setting to study the role of social interactions and users’ consensus in discussion forums as a source of collective intelligence for improving individual academic performance.

In on-campus education, social interactions are mainly mediated by collaboration and social abilities, which are highlighted as crucial competencies for life and work in the twenty-first century \cite{pllegrino}. Therefore, they are a large part of current educational endeavors \cite{11_McFarland,Rohrbeck2003,slavin2015cooperative,newman2002self,slavin2003cooperative,slavin2011instruction,pulgar2019physics,Pulgar19}. At the group level, social relations between peers are pivotal for social learning \cite{Henrich2015book,Henrich2001,pentland2015social,johnson1987learning,roger1994overview} and play a key role in academic outcomes \cite{Kassarnig2018,Blansky2013,Sacerdote2011,Davies2018,10_Biancani,pulgar2021student,candia2019strategic}. At the individual level, students’ positions in their social networks and their academic performance are significant and positively correlated \cite{Baldwin1997,Caprara2000,Mouw2006,8_Bruun,Gasevic2013,Blansky2013,Ivaniushina2018,Stadtfeld2019,Kassarnig2018}.

In contrast, in online education, social interactions and collaboration take a different form and occur mainly through computer-mediated communication \cite{Traxler2018}. Hence, social interactions respond to different demands linked to the features of digital technology. During emergency remote teaching, students have reported consistently poor and unsatisfactory interactions between peers and with the instructor \cite{almahasees2021faculty,coman2020online,sepulveda2020online,mseleku2020literature,van2020teachers}, which may contribute to the perception that synchronous activities are more effective \cite{nguyen2021insights,aristovnik2020impacts}. Therefore, it is relevant to study whether and how collaborative spaces provided by the LMS can contribute to the educational process.

Numerous studies conceptualize online participation using the frequency of posts in discussion forums (\cite{Romiszowski2004,Vonderwell2005} and reading time \cite{Hrastinski2009,Wise2013}. However, this approach is incomplete because the frequency of actions in discussion forums ignores interactions between people and emergent collective effects \cite{barabasi2003linked}. A more comprehensive approach for understanding participation and engagement lies in social network analysis \cite{Garton97}, where interactions are observed among students who jointly participate in forums \cite{Traxler2018}. Small-scale research literature has found positive relationships between students’ participation in online courses and academic success \cite{Morris2005,Traxler2018,Dawson2008}. However, these results depend on students’ readiness to deal with online demands \cite{Kebritchi2017} and their academic orientations \cite{dawson2010}.

Here, we conduct a large-scale study monitoring students’ interactions in discussion forums at a large university in Chile. We use tools from social network analysis to quantify students’ behavior in discussion forums and its effects on academic performance. Our motivation is based on the participants’ ability to produce, read, reply to, and value content, generating an emergent consensus on what information is valuable and what is not for their academic contexts. Thus, we aim to explore the role of discussion forums as a source of collective intelligence to improve individual academic performance.

\subsection*{Online learning}
The effectiveness of online education has been widely studied over the last decades. An increasing number of institutions are adopting its principles, particularly in higher education, because it facilitates access to a broader audience of learners while reducing restrictions in access to information \cite{panigrahi,Kebritchi2017}. Online education provides learners with the flexibility and autonomy to engage with course content according to their own pace and disposition. However, designing online courses involves several challenges for successful implementation and adoption. In addition to the lack of human interaction in learning as one of the limitations of online learning \cite{graham}, researchers have identified variables linked with learners’ expectations and skills that mediate their cognitive and social engagement in a course \cite{panigrahi,Kebritchi2017} as well as curriculum design and teaching strategies \cite{Nazir2015,hu2012}. Both elements are essential to foster students’ engagement and participation in online courses, a variable shown to be related to academic success \cite{Traxler2018,Morris2005,pye2015}. This section explores the different individual- and course-level variables that contribute to students’ participation and success in online courses.

Researchers have found that learners’ expectations, readiness, sense of belonging, identity, self-efficacy, and participation are critical for satisfaction and success in online courses \cite{Kebritchi2017}. First, having appropriate learning expectations might increase adaptability and optimize learning opportunities \cite{Li2008,Lyons2004,alraimi2015,Bhattacherjee}. Second, the flexibility and autonomy embedded in online courses require students to have the ability to manage their own time \cite{Hill2002,Roper2007}. Within the readiness framework, scholars have studied issues such as the level of technical skills for the use of computers and the internet \cite{Peng2006}. Extending the latter, Hung and colleagues \cite{Hung2010} suggested a five-dimensional model of readiness that includes self-directed learning, motivation for learning, computer and internet self-efficacy, online communication self-efficacy, and learner control. Self-efficacy is defined as individuals’ judgment of their own ability to perform different types of tasks \cite{bandura}. Consequently, Information and Communications Technologies (ICT) self-efficacy is related to learners’ confidence in utilizing computers and the internet and conveying ideas via the available digital technologies. These subdimensions of readiness stress the importance of autonomy and motivation toward learning as well as the ability to deal with and manipulate ICTs and their forms of communication to effectively navigate the advantages and challenges of online learning \cite{Kebritchi2017}.

Furthermore, the lack of human connection in online courses might lead learners to feel isolated from their classroom peers, thus affecting their sense of identity and further academic expectations and learning \cite{Koole2014}. Successful online courses encourage collective rather than individualistic cultural norms as the former foster familiarity and trust among members. This consequently facilitates a sense of belonging and the intention to share knowledge \cite{zhao2012}, thus improving the quality of discussions \cite{shih2014}. Therefore, it is vital to encourage in learners a sense of belonging within online learning environments, enabling the further development of academic goals and purpose \cite{Koole2014,Lapadar2007}. For instance, collaborative class participation in MOOCs (massive open online courses) has been shown to foster further participation and reduce desertion \cite{Ferschke}.

In connection with learners’ identity, participation in different online learning formats is another fundamental phenomenon discussed in the literature. Researchers recognize two forms of participation in online courses: participation as interactions between peers and instructors via written comments and responses \cite{Romiszowski2004,Vonderwell2005} and participation by reading these interactions more than writing them \cite{Hrastinski2009,Wise2013}. Within this line of research, scholars have challenged the dichotomy between active participation (i.e., writing in online courses) and passive participation (i.e., reading online comments) given that the latter demands active engagement with and reflection on the content \cite{Kebritchi2017,Hrastinski2009}. A study conducted by Wise and colleagues \cite{Wise2013} found that approximately three-quarters of students’ time is spent reading and observing written forms of interactions in online courses. When measuring participation and its effects on learning, researchers should focus on both forms of participation, the frequency of written posts and the time spent observing and reading written messages in online forums. A study conducted at the University of Georgia found that time spent reading written messages, was a predictor of final grades \cite{Morris2005}.

In addition to learners’ preparation to deal with online courses, their experience is mediated by the curriculum design and course structure. Researchers have reported a myriad of recommendations for increasing students’ participation and engagement in online courses. For instance, effective online courses should be centered on students \cite{Chametzky2014,Luyt2013} with emphasis on collaboration \cite{Niess2013,gupta2013,Tsai2013}. Additionally, communication in online courses requires guidelines for developing a climate of psychological safety for open communication, which has been shown to foster students’ participation over time \cite{zhao2012}. Such considerations directly contribute to learners’ sense of identity and belonging because students have the opportunity to work in their zones of proximal development alongside others. These principles do not differ from the social principles widely predicated in face-to-face education, yet their adoption in online formats requires further consideration of technical characteristics. Accordingly, the use of diverse digital technologies might enable further motivation to engage in the cognitive and social processes designed in online courses \cite{panigrahi}. The research-based recommendations for developing content in an online course include the integration of technology within a mix of collaborative and reflective activities and with clear performance expectations and assessment \cite{Niess2013}.

\subsection*{Online learners' participation}
Learners’ participation in online learning environments has been studied following quantitative (e.g., frequency of posts \cite{Morris2005}) and qualitative perspectives (e.g., content analysis \cite{deweber}). More recently, researchers have turned to social network analysis (SNA) in an attempt to comprehend such participation using relational measurements such as students’ network centrality and their structural roles in the network \cite{Traxler2018,Garton97}.

For instance, Traxler and colleagues \cite{Traxler2018} explored participation networks in consecutive online learning courses, conceptualizing network ties as joint interactions by posting on the same online forums; these authors found significant correlations between network centrality and grades. Because network centrality provides information on actors’ social integration in the network, Dawson \cite{Dawson2008} studied the relationship between network centrality and reported levels of the sense of community in the context of computer-mediated communication (CMC) and found positive correlations for closeness and degree centrality and a negative correlation for betweenness and sense of community. A subsequent study by Dawson showed network and performance-level differences between high- and low-performing students in online forums. In detail, high-performing students enjoyed more extensive networks whose members had high average grades compared to their low-performing counterparts \cite{dawson2010}. Additionally, and consistent with the expert versus novice literature, the content analysis revealed that high-performing networks were centered on conceptual content ideas, while in the low-performing network, the questions and comments focused on fact-based questions.

The above studies utilize SNA principles to conceptualize participation as aggregated interaction with forum posts. This relational perspective enables further examination of its connection with the academic success and sociodemographic features of students. In this work, we take an additional step in understanding the effects of online participation on academic success by conducting a large-scale study of all undergraduate students in their first and third years at a university in Chile. We map their coparticipant network and content exposure in discussion forums to evaluate its impact on academic performance.

\section*{\label{sec:resultados}Results}
\subsubsection*{Data}
Are discussion forums a source of collective intelligence that improves students’ academic performance? We investigated the role of network centrality metrics in discussion forums, which  capture forum and individual features related to the amount and quality of the information to which students were exposed, on the individual final GPA for all 7,528 students in their first year (n=3,585) and third year (n=3,943) in a large Chilean university. We only considered students who entered university via regular admission and enrolled in daytime modality degree programs, mainly because alternative admission methods respond to different students’ profiles and different institutional dynamics. Finally, we filtered out all students whose final GPA in semester 1-2021 was under 4.0, the minimum pass mark in Chile, because they had incomplete information in the university records. Thus, after the data filtering process, we analyzed three groups of students.

\begin{table}[h!]
\centering
\tiny
\caption{Data description for  hierarchical models.}
\label{data_table}
\begin{tabular}{p{0.15\linewidth} | p{0.15\linewidth}| p{0.3\linewidth}| p{0.3\linewidth}}
Variable &Type &Description & Note \\
\hline
\hline
Ave. GPA 2021-1 & Dependent variable &
  Corresponds to the final GPA of each student after the first class semester (July 2020).
  &
We run three models where we study first-year students, third-year students enrolled in 2018 (no failed subjects), and third-year students enrolled before 2018 (failed one or more subjects). Collected from the university records.
\\
\hline
Sex &
  Fixed effect &
  Student’s sex registered by the university (female or male). &
  Collected from the university records. 
\\
N. Credits &
  Fixed effect &
  Students’ total number of academic hours enrolled. &
The sum of the in-class time that each subject requires. Collected from the university records.
\\
Forum Writings &
  Fixed effect &
  Total number of posts written by each student. &
  All the written posts are collected for each student from the CANVAS LMS platform.
\\
Forum Readings &
  Fixed effect &
  Total number of posts read by each students. &
  All the read posts are collected for each student from the CANVAS LMS platform.
\\
High School GPA &
  Fixed effect &
  Student's average high school grade. &
  Collected from the university records. 
\\
Age &
  Fixed effect &
  Student's age in July 2020. &
  Collected from the university records.
\\
  \hline
Enrolled Degree Program &
  Random effect &
  Each degree program has different structures and number of classes, among others. We control for this using random effects. &
  Each degree program has its own intercept. Collected from the university records.
\\
High School &
  Random effect &
  Students’ high school is a proxy for income and cultural capital. We control for this using random effects combined with commune &
  Each high school from each commune has its own intercept. Collected from the university records.
\\
Commune (geographical) &
  Random effect &
  Students’ origin commune is a proxy for income and cultural capital. We control for this using random effects combined with high school. &
  Each high school from each commune has its own intercept. Collected from the university records.
\\
   \hline
Entrance Path &
  Filtering variable &
  Students can enter the university in several ways, such as sports scholarships, foreign admission, and highly vulnerable background, among others. We consider only enrolled students enrolled by the regular path, corresponding to more than $90\%$ of all students. &
  Collected from the university records.
\\
Cohort &
  Filtering variable &
  Students' cohort. &
 Used to filter out students taking classes from other cohorts. Collected from the university records.
\\
Entrance year &
Filtering variable
   &
  The year when students entered the university for the first time. &
  Used to filter out students who were previously enrolled in other degree programs. Collected from the university records.
\end{tabular}
\end{table}

\begin{itemize}
    \item Group I: First-year students who entered the university for the first time in 2020 (n=3,585).
	\item Group II: Third-year students who entered the university for the first time in 2018 (i.e., students who had not failed any subject) (n=2,296).
	\item Group III: Third-year students who entered the university for the first time before 2018 (i.e., students who had failed one or more subjects) (n=1,647).

\end{itemize}
We measured students’ academic performance–our dependent variable–as the average GPA at the end of the first semester of 2020. We controlled for several students’ confounders ranging from sociodemographic features to individual online behavior (Table \ref{data_table}). Sociodemographic variables were gathered from the university records, and online behavior variables were gathered from the databases of the Learning Management System (LMS) adopted by the University, CANVAS (for more details of CANVAS data extraction, see supplementary methods \ref{SM_Data}).

\subsubsection*{Co-participation in discussion forums}
\begin{figure}[t]
\centering
\includegraphics[width=0.99\linewidth]{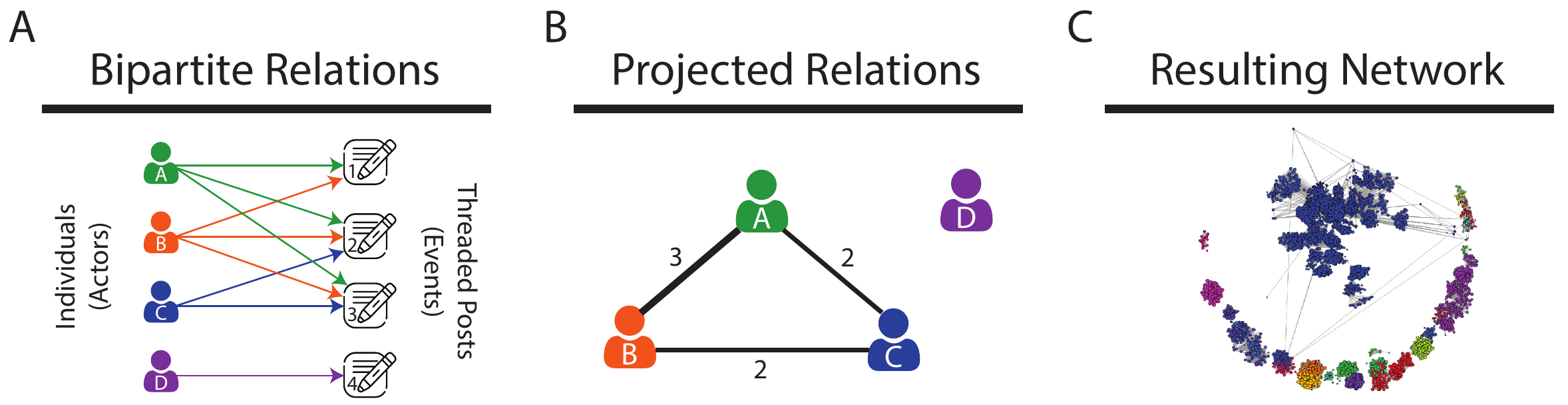}
\caption{Graphic depiction of the relationship mapping in CANVAS classroom forums.  \textbf{A)} Bipartite relations. There are two groups of nodes individuals or actors and threaded posts or events. Individuals can participate in different threaded posts. Arrows indicate the actors that participate in each threaded post\textbf{B)}. Individuals relate to each other if they participate in the same threaded post; for instance, actors A and B participate together in threaded posts 1, 2, and 3. Therefore, they have participated in three different discussion forums. Note that actor D only participates in threaded post 4, and no one else participates there. Therefore, actor D has no connections.  \textbf{C)} The resulting network of read threaded posts for first-year students; colors represent connected components. Note that the network also considers instructors and teaching assistants because they are relevant actors in the content diffusion process. See Figure \ref{redes_SM} for writing posts and third-year networks.}
\label{esque1}
\end{figure}

To assess the role of discussion forums as a valuable source of information for students, we extract and build three features from them at the student level (see Supplementary Methods \ref{SM_Data}). i) The relational network (Fig. \ref{esque1}) at the individual level, where students are connected if they participated significantly in different discussion forums. Then, we computed different network centrality measures (Table \ref{network_metrics}). ii) The number of read, written, and liked posts at the forum and individual levels to validate our network centrality measure and to control for LMS individual activity. iii) The text content of each student’s post for further analysis using natural language processing.

We elicit a network of co-participation for both first- and third-year students (Fig. \ref{esque1}) to build a bipartite network that relates actors (students, teachers, and teaching assistants) and events (LMS CANVAS threaded forums posts) (Fig. \ref{esque1}A). Given that we are interested in individual participation, we project the bipartite network on the actor dimension, linking individuals if they participated (wrote or read) in the same threaded post (Fig. \ref{esque1}B). Hence, we obtain an individual network for each written and read post (Fig. \ref{esque1}C) in each year of academic training.

We note that the mapped coparticipant network using only written posts captures the information of fewer students than reading posts, and the role of discussion forums as a source of collective intelligence relies on students consuming and being exposed to content created by others. Therefore, we choose to work with the read posts co-participation network (for more details about the network mapping and filtering process, see Methods, Supplementary Methods \ref{sup_filtering} and Fig. \ref{redes_SM}).

\subsubsection*{Network centrality to quantify the exposure to collective intelligence}

\begin{figure}[!t]
\centering
\includegraphics[width=0.99\linewidth]{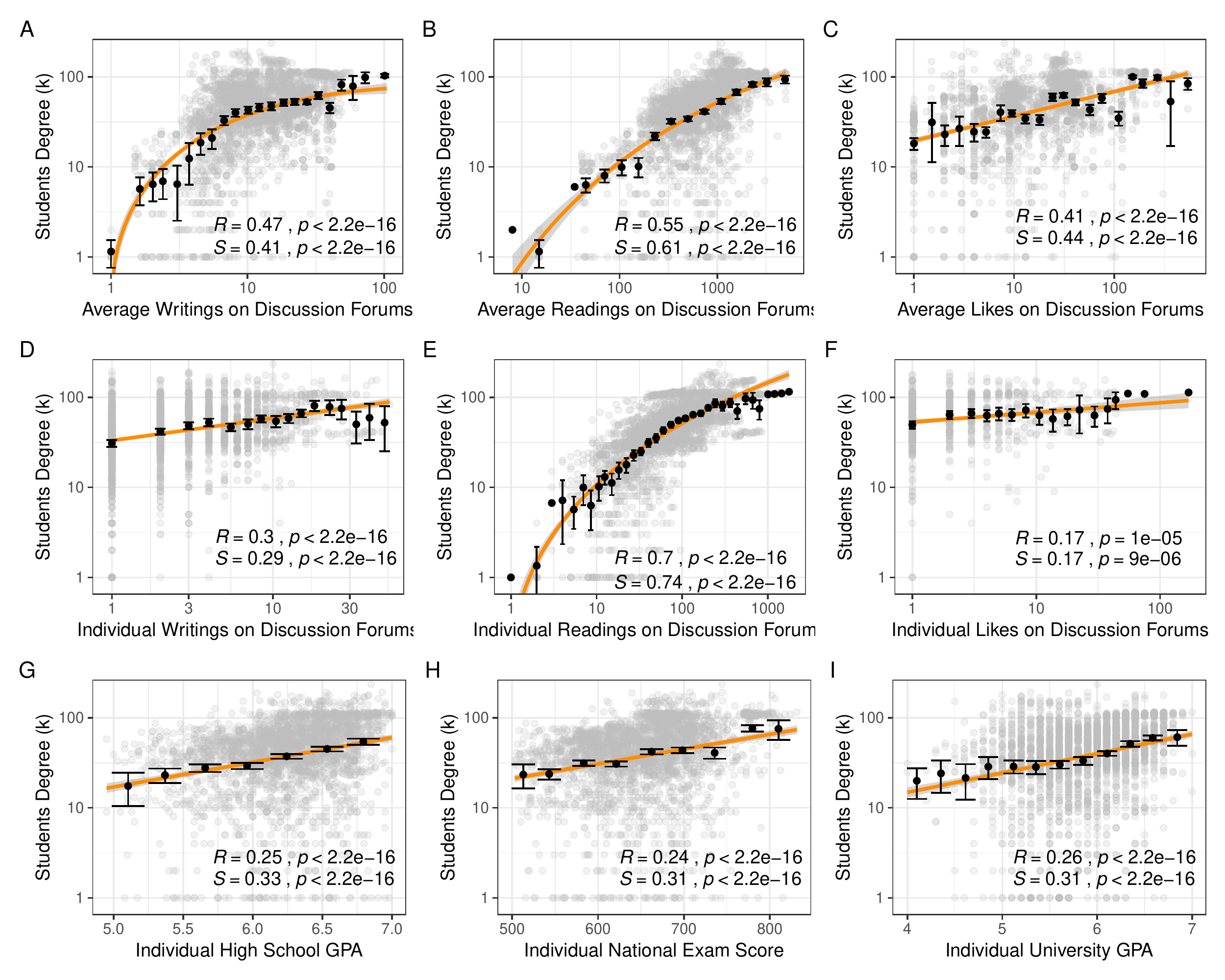}
\caption{Node degree captures forums and individual features for first-year students. R and S represent the Pearson and Spearman correlations, respectively, and p is their associated p value. We replicate these results for third-year students enrolled in 2018 (Group II, Fig. \ref{degree_SM1}) and third-year students enrolled before 2018 (Group 3, Fig. \ref{degree_SM2}). For the regression models see Table \ref{degree_table}. Note that the models do not exhibit a considerable variance inflation factor (Tables \ref{vif_degree_1}, \ref{vif_degree_2}, and \ref{vif_degree_3}))}
\label{node_degree}
\end{figure} 

The node degree captures both the features of both students and discussion forums. We focused on the network node (student) degree because more relations occur in important or relevant posts \cite{kleinberg1999hubs}, and the network degree offers a simple measure that quantifies the activity level in discussion forums (more reading and writing leads to a higher network degree). Moreover, the node degree also accounts for posts’ importance because important posts will be read, replied to and liked by more students, leading to a higher network node degree. These two characteristics motivate the node degree as a proxy for collective intelligence exposure \cite{levy1997collective}. We validate this idea by computing Pearson and Spearman correlations and running linear regression models (see Table \ref{degree_table}, and Figures \ref{node_degree}, \ref{degree_SM1}, and \ref{degree_SM2}).

Figure \ref{node_degree} shows the Pearson (R) and Spearman (S) correlations and their respective p values for the node degree and different forum features (average of written/read/liked posts across all discussion forums in which students participated), individual online behavior (the total number of writings/readings/likes of each student in LMS Canvas), and individual academic performance (high school GPA, application score to their degree program, and university GPA at the end of their first semester in 2020). We find positive and significant correlations for all these variables for first-year students, suggesting that the node degree is a strong measure of exposure to important posts, individual activity on discussion forums, and previous and current academic performance. We find similar results for third-year students from group II (Fig. \ref{degree_SM1}) and group III (Fi. \ref{degree_SM2}). Finally, we reinforce these results using linear regression models to explain the node degree in each analyzed group (Table \ref{degree_table}). We find that the node degree variance explained in each group is $68\%$ (Group I), $63\%$ (Group II), and $72\%$ (Group III). Our results survive after controlling for sex, the number of inscribed credits (proportional to the academic hours required by each class), students’ age, fixed effects per degree program, high school, family income, the ranking of the enrolled degree program, and the length (in semesters) of the enrolled degree program. Considering all control variables, the explained variance rises to $81\%$ (Group I), $87\%$ (Group II), and $85\%$ (Group III). We note that the variance inflation factors are all at low and moderate levels (see Tables \ref{vif_degree_1}, \ref{vif_degree_2}, and \ref{vif_degree_3}), indicating no multicollinearity problems.

We hypothesize that the effect of co-participation will be different for each group given their particular features. For instance, first-year students (group I) do not know each other; therefore, they do not have any other communication channel but the university LMS. Third-year students enrolled in 2018 (group II) already knew each other for two years. They probably had alternative communication channels; however, the discussion forum features (creation, replying, reading, and valuation of posts) are attractive enough for collective intelligence to emerge. Finally, third-year students enrolled before 2018 (group III) could already know each other, although they may not because they could belong to different previous cohorts. Nevertheless, their self-regulation and autonomous engagement with the course content are probably lower than those of group II given that these students have failed at least one subject. Therefore, an effect of collective intelligence on academic performance in this group is less likely.

\subsubsection*{Discussion forums as a source of collective intelligence}

\begin{figure}[h!]
\centering
\includegraphics[width=0.99\linewidth]{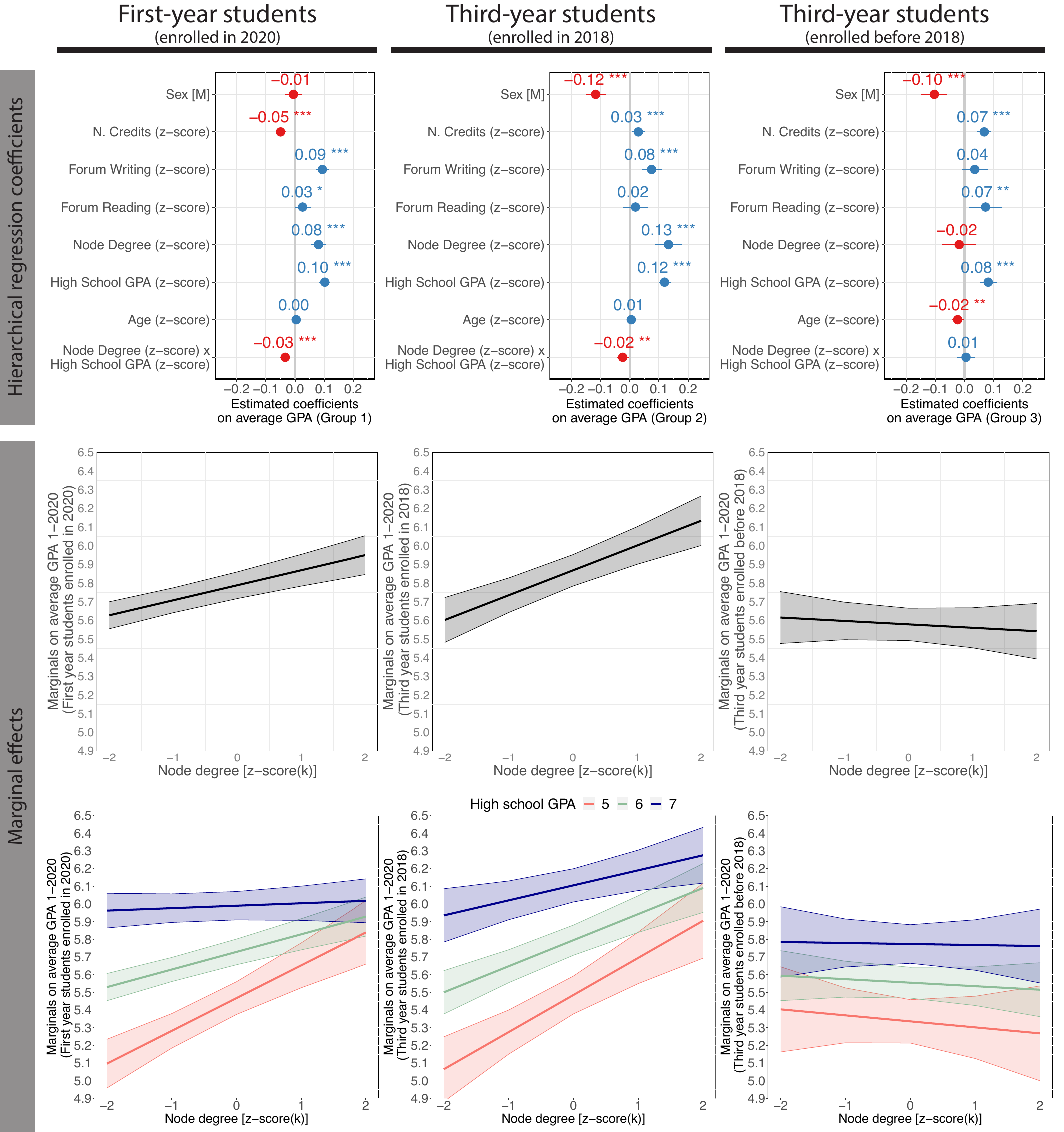}
\caption{: Hierarchical regression models for the average GPA obtained by three samples of students at the end of the first semester of 2020. The first column refers to first-year students enrolled in 2020. The second column refers to third-year students enrolled in 2018, i.e., students who have not failed a subject. The third column refers to third-year students enrolled before 2018, i.e., students taking overdue subjects. We show the standardized regression coefficients (first row) to compare the effects between different independent variables. The second row depicts the marginal effects on the standardized number of students’ connections in the forum coexposure network built from online class forum readings, quantified as the node degree (k). The third row depicts the same marginal effects but for different high school GPAs. (See Table \ref{main_table} for more details; Fig. \ref{main_sm} for nonstandardized regression models, and Fig. \ref{SM_diag_plot} for the diagonal plot of each model).}
\label{fig1M}
\end{figure} 

We bring our data to hierarchical models according to the following specification:

\begin{equation}
    GPA_{i}=\beta_{0i[jl]} + \beta_{1i[jl]}K_i + \beta_{ni}X_{ni},
\end{equation}
where $GPA_i$, our dependent variable, is the final GPA for each student $i$ at the end of the first semester of 2020, immediately after the emergency remote teaching modality started. The motivation to use hierarchical models is that the final GPA has different levels and increases at different rates with the node degree ($K_i$) depending on the student’s enrolled degree program j and the student’s former high school $l$. Therefore, $\beta_{0i}[j]$  represents the random intercept for each enrolled degree program ($j$) and each commune-high school combination (l), $K_i$ is the student $i$ node degree, and $\beta_{1i}[jl]$  represents the random slope of node degree for each degree program and each high school. We observe that the effect of node degree varies with the knowledge area (academic department): social sciences has the strongest effect and engineering has the weakest effect (Fig. \ref{fig4M}). Finally, $X_{ni}$ represents an n-dimensional coefficient vector for each of the n control variables for each student $i$ (see Table \ref{data_table} for a description of the included control variables).

The results (Fig. \ref{fig1M} first row) show that the node degree, $K_i$, has a significant and positive effect on students’ final GPA for first-year students and third-year students enrolled in 2018 beyond and above controlling for the number of read and written posts. Moreover, we observe a negative and significant interaction term between the node degree and high school GPA for the same groups. We note that the node degree effects are not significant for third-year students enrolled before 2018. These results suggest that networked participation has extra value in capturing students’ behavior of forum participation (see Table \ref{main_table} for the regression models, Fig. \ref{main_sm} for the model without standardizing the node degree, and Fig. \ref{SM_diag_plot} for the diagonal plot of the model that graphically shows a noticeable correspondence between the predicted and observed final GPA).

Interpreting interaction coefficients for two continuous variables can be challenging; hence, we plot the unconditional marginal effects of node degree on final GPA (Fig.\ref{fig1M} second row). We observe that students’ node degree has a global positive effect on the final GPA even after considering the negative interaction and including all the available control variables.

However, what does the negative effect of the interaction term mean? To disentangle the relationship between high-school GPA and node degree, we plot the marginal effects of the node degree on final GPA conditioned on high school GPA. We observe for first-year students and third-year students enrolled in 2018 that the positive effect of the node degree is steeper for low high-school GPA students (red color, p value $<0.01$) compared to students with a high high-school GPA (blue color). Note that for high high-school GPA (blue line) first-year students, the effect is practically nonsignificant. Finally, there is no significant effect between the node degree and final GPA for third-year students enrolled before 2018 (Fig.  \ref{fig1M} third column).

\subsubsection*{Robustness checks}
As robustness checks, we run the model without the interaction term and obtain strong positive effects for the node degree on final GPA for both first-year students and third-year students enrolled in 2018 and no effect for third-year students enrolled before 2018 (Fig. \ref{SM_no_interaction}). Nevertheless, an ANOVA shows that the interaction term between the node degree and high school GPA is significant for improving the performance of the model (p value $<0.05$) for both first-year students and third-year students enrolled in 2018 (Table \ref{anova_main}).

To show that our results do not depend on the specific measure of prior individual ability, we replicate our results using the national exam average score (PSU score) for application to higher education (Fig. \ref{Res_SM3}) instead of high-school GPA. Our results suggest that the main finding is robust for a different proxy of students’ ability.

We explore whether the results are an artifact of the chosen centrality measure. We replicate our analysis by replacing the node degree with the PageRank centrality (Table \ref{network_metrics}), obtaining the same results (Fig. \ref{SM_PageRank}). We also explore two other qualitatively different network centrality measures, the clustering coefficient and Burt’s constraint. Using the clustering coefficient, we test whether access to information within highly connected neighborhoods (high clustering coefficient, Table \ref{network_metrics}) has a role in the final GPA. Figure \ref{fig3M} shows that co-participating in forum posts in close groups has a negative, although small, effect on final GPA for first-year students and for third-year students enrolled in 2018. For third-year students enrolled before 2018, we do not observe a significant effect. We provide more evidence on this point by replicating these results using Burt’s network constraint, which quantifies access to redundant information (Table \ref{network_metrics}). We find no significant effect of Burt’s constraint on final GPA (Fig.  \ref{SM_constraint}).

\subsubsection*{Content exposure in discussion forums}
We have replicated our main findings using alternative measures of prior individual ability (high school GPA and national examination score) and network centrality measures (node degree and PageRank). Additionally, we found divergent validity on different centrality measures, which quantifies access to redundant information (cluster coefficient and Burt’s constraint centrality). Now, we ask why the effect of node degree on final GPA is steeper for disadvantaged (low high-school GPA) students.

\begin{figure}[t!]
\centering
\includegraphics[width=0.99\linewidth]{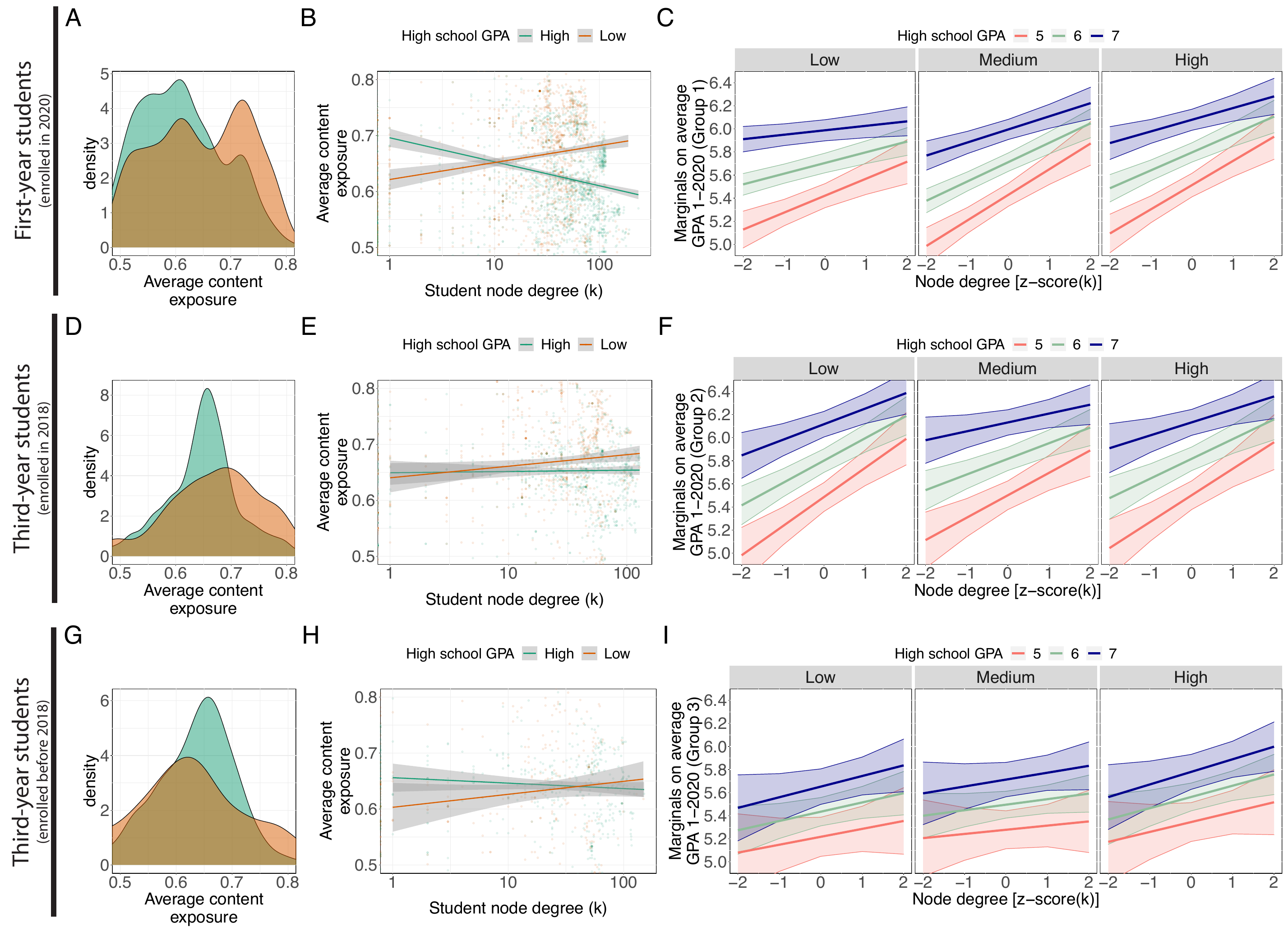}
\caption{First-year students who are exposed to posts with high content intensity benefit most from discussion forums (see Table \ref{bottom_top1}). The first column shows the distribution of the individual average content exposure for high (green) and low (orange) GPA students. The second column shows the relationship between the individual node degree and average content exposure for high (green) and low (orange) GPA students. The third column shows the marginal effects for the main model adding an interaction term between the node degree and average content exposure (see Tables \ref{bottom_top1}, \ref{bottom_top2}, and \ref{bottom_top3} for more details). Low, medium, and high refer to the average content exposure.}
\label{fig5M}
\end{figure} 

Here, we address the fact that there is significant variability in the type of texts written in each forum, ranging from subject-related content to social-related texts. We use a neural network approach to quantify the content intensity of each forum post text (see Methods Forum Posts Semantics). We define content intensity as the cosine similarity between each written text and the content that each department taught. For instance, we characterize the computer science department with all the words (tokens) in the course names. A random sample of the tokens for the Department of Computer Science is "\textit{cybersecurity, computing, commerce, technologies, society, optimization, algorithm, logic, introduction, structure}". Then, we compare the similarity between these tokens and each text post written in a computer science course. To create the intuition, we provide two text examples. The first is a text post with high content intensity (cosine similarity of $0.89$):

\begin{quote}
\textit{\textbf{Defining what is Information Technology.} Computer technology is related to how data are managed and handled; within an industry, it can be understood as all systems that store, process, transmit, convert, copy or receive electronic information. An example of what computer technology would not be is using rudimentary methods such as managing shifts through books, storing employee data on paper, etc.}
\end{quote}

An example of a post with low content intensity (cosine similarity of $0.58$) is the following: 

\begin{quote}
\textit{\textbf{Presentation of weekly themes.} The thing is you have to be committed to adding value, not to contributing hours.}
\end{quote}

We observe that the first post is related to computer science content, and the second one is more related to social aspects of the course (for more details, see Method section Forum Posts Semantics).

Thus, we ask whether the steeper effect of the node degree network centrality on final GPA for disadvantaged students depends on the type of forum post text in which they participate, quantified as the content intensity. The first column in Figure \ref{fig5M} shows the density of the individual average content exposure, quantified as the average of content intensity among all the post texts in which each student participates per high school GPA (green density corresponds to high GPA and orange density to low GPA). The average content exposure is similar for each cohort, $0.62\pm0.08$ for first-year students enrolled in 2020, 0.65±0.09 for third-year students enrolled in 2018, and $0.61\pm0.10$ for third-year students enrolled before 2018. The second column in Figure \ref{fig5M} shows the relationship between the student node degree and the average content exposure conditioned on high school GPA. We find a positive relation for low high-school GPA students (orange line) and a negative relation for high high-school GPA students (green). This effect is the strongest in first-year students and is mild for third-year students enrolled in 2018. We do not find a significant effect for third-year students enrolled before 2018. Thus, our results suggest that low high-school GPA students whose node degree in discussion forums is high interact in more content-intensive forum text posts than low high-school GPA students. We test this idea, replicating our main result and adding an interaction term between content intensity and node degree (third column Figure \ref{fig5M}). We find that first-year students who are exposed to high content-intensity posts (high panel) show the strongest relation between the node degree and final university GPA (for more details, see Tables \ref{bottom_top1}, \ref{bottom_top2}, and \ref{bottom_top3}).

\section*{\label{sec:disc}Discussion}
The reality of emergency remote teaching can be particularly challenging for first-year university students due to the impossibility of developing in-person relationships of friendship and collaboration, which are pivotal to accessing information and for emotional and academic support. Therefore, access to a source of collective intelligence that enables cooperative and consensus dynamics between students can enrich their learning experiences \cite{levy1997collective}.

Discussion forums are driven by collective intelligence, which allows people to write, read and value text posts, thus promoting cooperation and consensus on which are valuable texts and which are not. The simplest metric that captures discussion forums’ dynamics is the node degree, defined as the number of significant students’ connections in the underlying co-participation network. Indeed, the node degree is positively and significantly associated with features from the discussion forums, individual online behavior, and previous and current academic performance (Figures \ref{node_degree}, \ref{degree_SM1}, and \ref{degree_SM2} and Table \ref{degree_table}). These results validate the idea that the node degree is a good proxy of exposure to collective intelligence. Our results show that a high node degree is positively and significantly associated with higher university GPA in first- and third-year students enrolled in 2018 (Fig. 3 and Table \ref{main_table}); these effects survive even after controlling for several confounder variables regarding individual online behavior and individual features. We replicate our results by interchanging node degree with an alternative network centrality, the PageRank (\ref{SM_PageRank}) and high school GPA by the score in the national exam for university application (\ref{Res_SM3}).

The positive effect of the node degree on final GPA is steeper for first-year students with a low high-school GPA (Fig. \ref{fig1M}) (or low university application scores, Fig. \ref{Res_SM3}). Presumably, these students entered higher education with deficiencies in learning content and skills to meet the demands of undergraduate courses. Thus, our results suggest that participation in discussion forums can serve as a source of collective intelligence, and exposure to content created on them helps disadvantaged first-year students to a greater extent. We hypothesize that access to relevant information regarding subject content leads to higher individual academic performance. Using natural language processing tools, we quantify the content intensity of each text in discussion forums, i.e., how related each text is to the content taught by the reader’s academic department. We find that first-year students with low high-school GPA are exposed, on average, to high content-intensive texts. When we replicate our main analysis separated by low, medium, and high content-intensity texts, we find positive and significant evidence for our hypothesis: disadvantaged first-year students benefit the most from discussion forums because they consume more content-intensive texts (Fig. \ref{fig5M} and Tables \ref{bottom_top1}, \ref{bottom_top2}, and \ref{bottom_top3}).

The forums and messages written by their peers regarding the course content become  alternative mechanisms to access information and build learning. It could be argued that forums—unlike messages shared in face-to-face contexts whose access is limited by particular social relationships in a given time and require the acknowledgment of students’ particular ignorance on a certain topic—facilitate access to information for all participants in the course in an anonymous fashion, which has been studied as a relevant mechanism for encouraging collaboration \cite{freeman2004student,song2015hidden,chester1998online}. Consequently, from the perspective of Vygotsky’s zone of proximal development (ZPD) \cite{vyg}, the forums not only allow the writing and discussion of the content to be learned but also expand the limits of possible learning to those who enter the first year in a situation of academic disadvantage. In this way, participation in discussion forums reduces the differences in performance between students of different academic conditions in the first year of university. This is also evidenced in third-year students enrolled in 2018 (second column Fig. \ref{fig1M}).

It is also interesting to recognize that first-year students who enter the university with better school grades and national exam scores do not benefit from participating in the forums (blue line in first column Fig. \ref{fig1M}). In line with this interpretation, students in such conditions may have skills, competencies, and support networks to access relevant information for success in higher education. This interpretation is also sustained by third-year students with good high school performance. Another potential explanation is that advantaged students trust others’ knowledge and contributions less \cite{tschannen2001collaboration}. Thus, advantaged students may have reputation incentives to write posts but no incentives to consume others’ content.

Finally, we found divergent validity in third-year students enrolled before 2018, i.e., those who took overdue subjects from the third year (Group III). Greater participation in discussion forums does not benefit their academic performance. This phenomenon could be explained by the fact that these students are repeating subjects and may not need forums to build learning since they may have previous experience in these contents and social networks outside the subject to improve their performance. Another potential explanation is that Group III could arguably be students with low self-efficacy, self-regulation, motivation, and/or engagement; therefore, they participate less in everything (which could also explain their failed subjects). If they participated and engaged in discussion forums, they could have spent considerable time in the forums as a means of access to the content, but without apparent success.

Thus, our findings suggest that the collective intelligence embedded in discussion forums can even level the field for disadvantaged students in a remote learning modality. Figures \ref{fig1M} and \ref{fig5M} show that disadvantaged students with the highest node degree can even increase their final university GPA to the level of advantaged students. This consequence can be pivotal for developing efficient new strategies for encouraging and facilitating learning in disadvantaged and vulnerable students.

\section*{\label{sec:Met}Methods}

\subsection*{\label{sec:networks}Mapping discussion forum's co-participations}
The bipartite approach is broadly used, for example, to model scientific collaboration networks \cite{hou2008structure,tomassini2007empirical}, product exports in economics \cite{hidalgo2007product}, relations between different species in ecology \cite{dormann2009indices,saavedra2009simple}, relations between diseases in medicine \cite{barabasi2011network}, and the similarity of degree programs in higher education \cite{candia2019higher}. More recently, some online education studies have used this approach to capture student interactions on forums \cite{Traxler2018}. Here, we adopt the same approach.

To establish the effect of class forums’ interactions, we elicit a network of co-participation for both first- and third-year students (Fig. \ref{esque1}). Individuals can participate (read or write) in any threaded post in their registered courses, defining a bipartite relation between individuals or actors (students, teachers, and teaching assistants) and threaded posts or events (Fig. \ref{esque1}A). It is possible to project bipartite relations in dimensions, actors and events. We are interested in online human co-participation; hence, we link individuals if they participate in the same threaded post (Fig. \ref{esque1}B).

The projected bipartite networks are built from thousands of readings or writings on CANVAS discussion forums, with content ranging from conversations related to subjects to topics related to social bonding and with different levels of participation for different actors. Therefore, we can reasonably argue that not all co-participation is equally significant. The most active individuals may be connected by some central structure underlying the entire network, which is the structure that we need to elicit. Thus, to unveil the backbone of the co-participation network, we filter out all the links that can be explained by a null model. To this aim, we compute the $\phi$ correlations between actors, and then we compute a t test for correlation significance using a p value of $0.05$. Finally, we discard all negative correlations and nonsignificant correlations, and we remove all loose nodes (Fig. \ref{esque1}C and Fig. \ref{redes_SM}; for more details about the filtering process, see Supplementary method \ref{sup_filtering}).

\subsection*{\label{sec:centralities}Centrality metrics}
To explore and understand the forums’ navigation patterns associated with academic success, we calculate different and alternative network centrality metrics from the co-participation network (Table \ref{network_metrics}), including node degree, PageRank, local clustering, and Burt’s constraint. We use these student-level centrality metrics in our regression models to study the relationship between students’ forum co-participation and their final GPA in the first semester of 2020.

We focus on the network node degree (Table \ref{network_metrics}) as our main centrality metric that captures behavioral information from students interacting in discussion forums. To state the robustness of our findings, we use PageRank centrality, which also accounts for the centrality of the network neighbors of each given student, i.e., a connection with a central student in the co-participation network is weighted more than a connection with a noncentral student. The clustering coefficient and Burt’s constraint are used to explore the effect of accessing redundant information in discussion forums.

\begin{table}[!ht]
\caption{Network metrics and their definitions. $w_{ij}$ is the number of links between students $i$ and $j$, and N is the total number of students.}
\label{network_metrics}
\centering
\scriptsize
\begin{tabular}{p{0.35\linewidth} | p{0.6\linewidth}}
Notation &
  Definition \\ \hline
$G(V_{i}, E)$ &
  Network G is the backbone of the forum posts' network. V represents the vertex (students, i) and E is the edges. \\
 &
   \\
$k(V_{i}) = \frac{1}{N}\sum_{j \neq i} w_{ji}$ &
  Node degree is a centrality metric that quantifies the number of connections that a student has in the forum co-exposure network. \\
 &
   \\
$P(V_{i}) = \frac{1-d}{N} + d  \sum_{j=1}^n \frac{w_{ij}P_j}{\sum_{k=1}^n w_{kj}}$ &
  PageRank is a centrality metric that quantifies the number of connections and the importance of those connections for each student in the forum co-exposure network; d represents a dumping factor. The inventors of Page-Rank \cite{page1999pagerank} recommend to set it as $d=0.85$.\\
 &
   \\
$C_i={\frac  {|\{e_{{jk}}:v_{j},v_{k}\in N_{i},e_{{jk}}\in E\}|}{k_{i}(k_{i}-1)}}$ &
  The clustering coefficient is defined as the proportion of the number of links between the neighbors of vertex $i$ divided by the number of links that could possibly exist between them. The more neighbors are connected between them, the higher the clustering is, leading to a higher probability to access redundant information.\\
 &
   \\
$C'_i=\sum_j w_{ij}[1-\sum_q \frac{w_{iq}}{\sum_{j_{i\neq j}} w_{ij}}w_{jq}], q\neq i,j$ &
  Burt's constraint for an undirected network is a centrality metric that quantifies the access to structural holes in the forum co-exposure network. The higher Burt's constraint is, the higher the probability of accessing redundant information.
  
\end{tabular}
\end{table}

\subsection*{\label{sec:data_ana}Regression models}
To evaluate the impact of co-participation in discussion forums on the students’ GPA obtained at the end of the first semester of 2020, we apply all our data to regression models. Considering that each degree program works differently and has different characteristics and that different high school contexts have different effects on university outcomes, we opt to use mixed (or hierarchical) regression models \cite{gelman2006data}. Hierarchical regressions allow us to model the variance of the estimators induced by each group, defined by students’ degree programs and secondary school.

The analysis is conducted at the student level. Therefore, the dependent variable is the GPA across all the courses registered by each student at the end of the first semester of 2020. After testing different specifications, the proposed models consider as significant control variables the fixed effects of gender, age, the number of credits enrolled, and the high school GPA (or the weighted score in the national exam, PSU). The models also include random effects in the intercepts (each of the categories in each group has an independent intercept) for the degree program and the student’s secondary school (Table \ref{data_table}).

Finally, we run the analysis for all students in the daytime mode who were enrolled through the regular admission path. We note that the analysis covers all students with an average grade greater than or equal to 4.0 (the minimum passing grade in Chile), and we discard all students who dropped out of their degree programs within the first semester of 2020 because it is impossible to compute a reliable average GPA for analysis. Thus, the final number of analyzed students is $2,648$ first-year students, $1,467$ third-year students enrolled for the first time at the university in 2018 (no-failed subjects), and $638$ third-year students enrolled for the first time at the university before 2018 (with at least one failed subject).

\subsection*{\label{sec:semantics}Forum posts semantics}
We use Word2Vec \cite{mikolov2013efficient,mikolov2013distributed}, a word embedding technique that enables us to represent words in a vector space–or semantic space–where the semantic relations between words are represented by the vectors’ directions. We use a pretrained semantic space \cite{cardellinoSBWCE} built from a corpus of more than 2 billion Spanish tokens (words, combinations of words, and chunks of words).

Thus, we characterize and represent each university’s academic department in a semantic space. Each token is represented as a vector in the semantic space. In particular, the difference between two vectors’ directions indicates the semantic dissimilarity of the words and is quantified as the cosine similarity between vectors ranging, theoretically, between $-1$ and $1$. In other words, parallel vectors (minimum direction difference) indicate maximum similarity (cosine similarity equal to $1$), and orthogonal vectors (maximum direction difference) indicate minimum similarity (cosine similarity equal to $0$).

We consider all the words that describe the names of the courses taught by each academic department. We filter out all the stop words, and then we lemmatize all the words. The resulting groups of words are considered a proxy for the subjects’ content that characterizes each academic department. Then, using the vectorial sum, we build a semantic representation for each academic department in the semantic space.

Next, we represent each post in the semantic space summing each post’s word vectorially. Thus, we can estimate how close (in terms of content) each post is to its corresponding academic department (the academic department where the subject is being taught) in the semantic space. We compute the cosine similarity between both vectors (discussion forum post and department), which results in a content intensity measure for each written text in online discussion forums \cite{garten2018dictionaries}.

Finally, we compute a student’s average content exposure by averaging the content intensity of each post that a certain student read.


\section*{Acknowledgments}
The research reported in this publication was supported by Unidad de Fortalecimiento Institucional of the Ministerio de Educación Chile, project InES 2018 UCO1808 Laboratorio de Innovación educativa basada en investigación para el fortalecimiento de los aprendizajes de ciencias básicas en la Universidad de Concepción. This research was partially supported by FONDEF ID19I10413. C.C. thanks Javier Pulgar for his thorough and insightful comments and suggestions.

\section*{Author contributions statement}
The authors confirm contribution to the paper as follows: study conception and design: C. C, and K.L.; data extraction and data cleaning: F.P. and C.C.; model creation and data analysis: C.C.; interpretation of results: C. C, A. M and C.B.; original draft preparation: C. C and A. M; manuscript review and editing: C. C, A. M and C.B.; project administration: K.L.; research supervision: C.B.

\section*{Competing interests} 
The authors declare no competing interests.

\pagebreak

\setcounter{page}{1}

\section*{\centering{Supplementary material}}

\beginsupplement


\section{Supplementary methods}

\subsection{Data Gathering}\label{SM_Data}

We gathered data on students from Canvas LMS, adopted by the whole university. We focused on the Canvas Data Portal service, which through its API, canvas-data-cli, provides the data associated with student's activity on the platform, in an orderly and structured way. In the study we use the \textit{requests} records, which provides information at the level of the specific action performed by the user together with its date-time. These records are commonly called log records. 

The data cleaning process of the log records involved remove duplicates and records without associated user or action. Then, using regular expressions, we performed the processing of the log's url (url encompasses information on the location of the action with the following structure: \textit{/courses/10/discussion/280}. This url refers to an action in the course '10' and the forum discussion '280'). We also processed the \textit{web\_application\_action} field that encompasses if the action is 'read' or 'write'. 

After cleaning and processing the log records, we use the tables the \textit{course\_dim}, \textit{user\_dim} and \textit{dicussion\_entry\_dim} of Canvas Data Portal in order to obtain the course (subject) ID associated to each action, the user ID who performs the action, and the text message associated with each action (reading or writing).

The table \textit{dicussion\_entry\_dim} contains the identifier of each message, identifier of the forum to which it belongs, the user ID who writes the text message, and an indicator if the user replies to a colleague or replies in the forum root. On the other hand, the table \textit{discussion\_topic\_dim} reports who created the forum, the forum topic, and the creation time, containing the associated course to which they belong. 

The text posts were cleaned of html tags. Thus, we discarded comments that were only images, empty posts, posts that were only pure html tags. Besides, discussion forums without any comments were not considered. 

For the semantic analysis of text posts, the messages were tokenized and lemmatized. We discarded very short messages, posts with pure emojis, text in non-spanish language, or messages without tokens contained in the word embedding model (Word2Vec). 

Using the \textit{user\_dim} and \textit{pseudonym\_dim} tables we obtained the institutional identifier of the student, which was essential for joining behavioral data on the Canvas LMS with institutional records.

The institutional records of each students are extracted from BI tool used by the University, as well as the names of the faculties, departments and subjects (where we did a manual curation), among many others.

Finally, the dataset used in this article encompass both behavioral data from Canvas LMS and institutional records.

\subsection{Network filtering process}\label{sup_filtering}

We use all the data from CANVAS of all the individuals (students, student assistants and teachers) who wrote or read a post in the forums associated with all the courses associated with any first or third year curriculum.

For each individual, the data set comprises a list of all the forums where they participated passively or actively, either reading or writing a text. From here, a list of pairs of co-participation (joint reads and / or writes) of individuals is generated for each forum, that is, we associate individuals if they both participate in the same forum. After obtaining a list for each forum, we construct a matrix of co-participations, $M$, that counts the times that each pair of people participate in the same forum, $M_{ij}$, that is, if individual 1 and individual 2 participated 4 times in forum A and 3 times in forum B, the entry of the matrix associated with both individuals is defined as $M_{12} = 4 + 3 = 7$. Furthermore, we define $M_i = \sum_{j}M_{ij}$ as the total number of shares of each individual. Next, a projection of a mode is created by linking all the actors that were connected to the same event node (Figure 1). A student who posts multiple times in a thread will have links of greater weight to the other participants of the thread in the actor's screening. Add up the weights of multiple links between two people (from posting in multiple common threads).

The procedure described is known as the bipartite network model, recurrently used to model situations in which both people ("actors") and some set of shared activities ("events") are of interest. This approach has been used, for example, to model scientific collaboration networks \cite{hou2008structure,tomassini2007empirical}, where scientists (actors) and articles (events) are the two types of nodes. Also, this approach has been used in economics \cite{hidalgo2007product}, ecology \cite{dormann2009indices,saavedra2009simple}, and education \cite{candia2019higher}. More recently, some online education studies have used it to capture student interactions on forums \cite{Traxler2018}. The bipartite analysis is developed in two stages. First, the network is created with two types of nodes (individuals and forums). The bipartisan approach gives us a loud and strongly interconnected picture of forum conversations. The social network generated from the bipartite methodology has more links and more connections than typical networks based on class surveys. So, given that these networks build from thousands of readings and posts, with content ranging from conversations related to courses to topics related to social bonding, it seems reasonable that not all interactions are equally important. The most active individuals may be connected by some central structure underlying the entire "noisy" network, and it is these kinds of structures that we need to reveal. Therefore, the following steps are performed:

\begin{itemize}
    \item Calculate the $\phi$ correlations between pairs of individuals as $\phi_{ij}=\frac{M_{ij}Z-M_{i}M_{j}}{\sqrt{M_{j}M_{i}(Z-M_{i})(Z-M _{j})}}$
    
    \item Discard all links with negative correlations. Negative correlations represent people that co-participate in forums in a lower proportion than expected at random.
    
    \item Calculate the $t$ statistic to evaluate whether the correlations are significantly different from zero with a significance threshold of p-value $<0.05$ as $t_{ij}=\phi _{ij}\frac{\sqrt{D-2}}{\sqrt{1-\phi _{ij}^{2}}}$.
    
    \item All loose nodes are discarded.
    
\end{itemize}

We map four participation networks based on the type of participation (reading or writing) and the courses enrolled (first or third year) (Figure 3). We note that of the 6,433 students who are enrolled in first-year courses, the network contains information on 4,901 students and capturing information on 117 assistant students and 208 teachers. In comparison, the network of co-participation (writing) for first-year students contains information on only 1050 students, 34 assistant students, and 104 teachers. On the other hand, the network of third-year courses, out of 4543 students enrolled in third-year courses, captures information from 3705 students, 129 assistants, and 175 teachers in reading-type participation and 944 students, 28 assistants 90 teachers in participation writing type.

\begin{figure}[h!]
\centering
\includegraphics[width=0.8\linewidth]{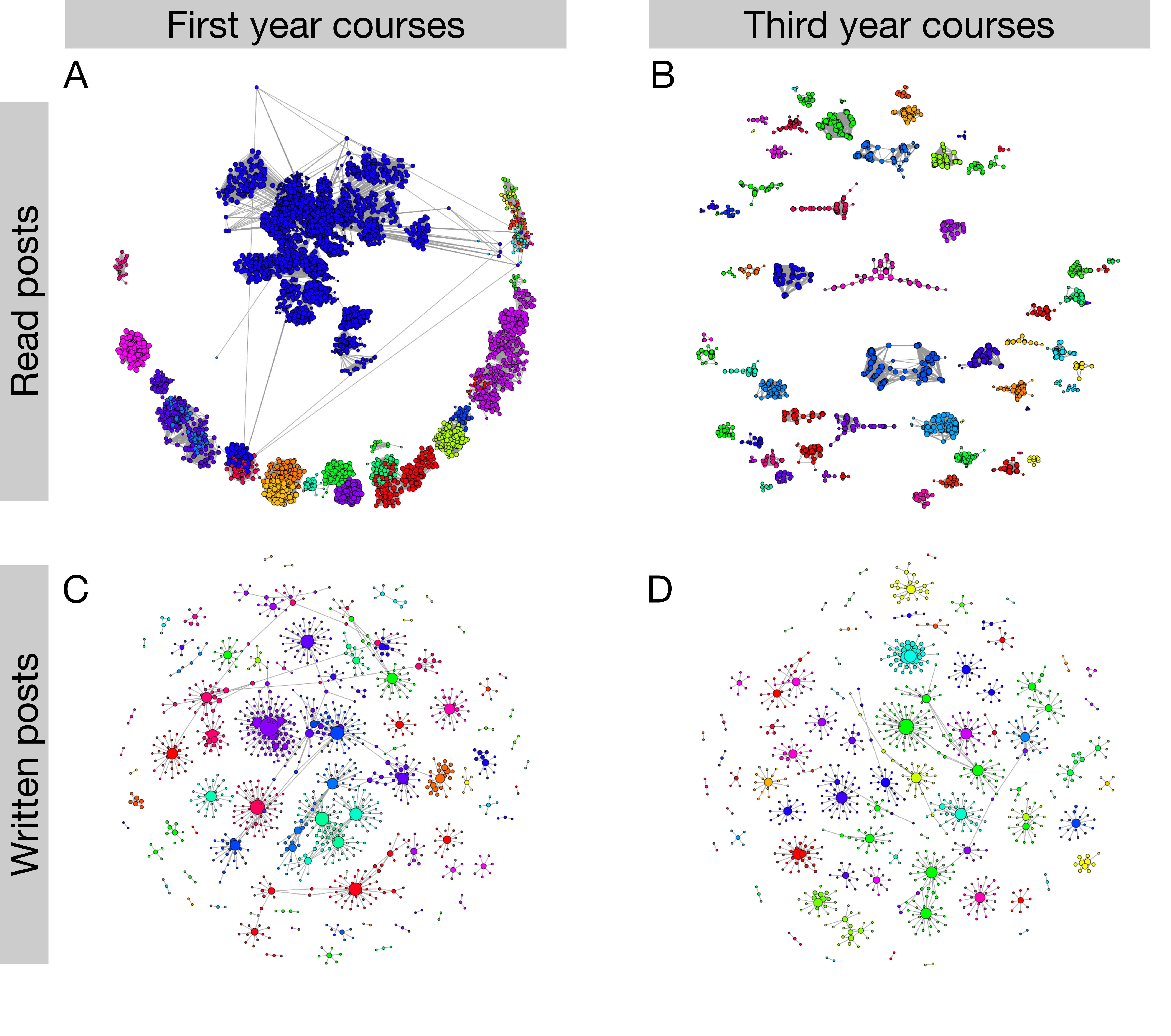}
\caption{Participation networks in online forums. A) Reading-type participation network among students enrolled in first-year courses. B) Reading-type participation network among students enrolled in third-year courses. C) Network of writing-type participation among students enrolled in first-year courses. D) Network of writing-type participation among students enrolled in third-year courses. Colors represent connected components, that is, people who have at least one path through other people connecting them within the network.}
\label{redes_SM}
\end{figure} 

\clearpage
\pagebreak

\subsection{Filtered network correlations}\label{sup_cetralities}

\begin{figure}[h!]
\centering
\includegraphics[width=0.9\linewidth]{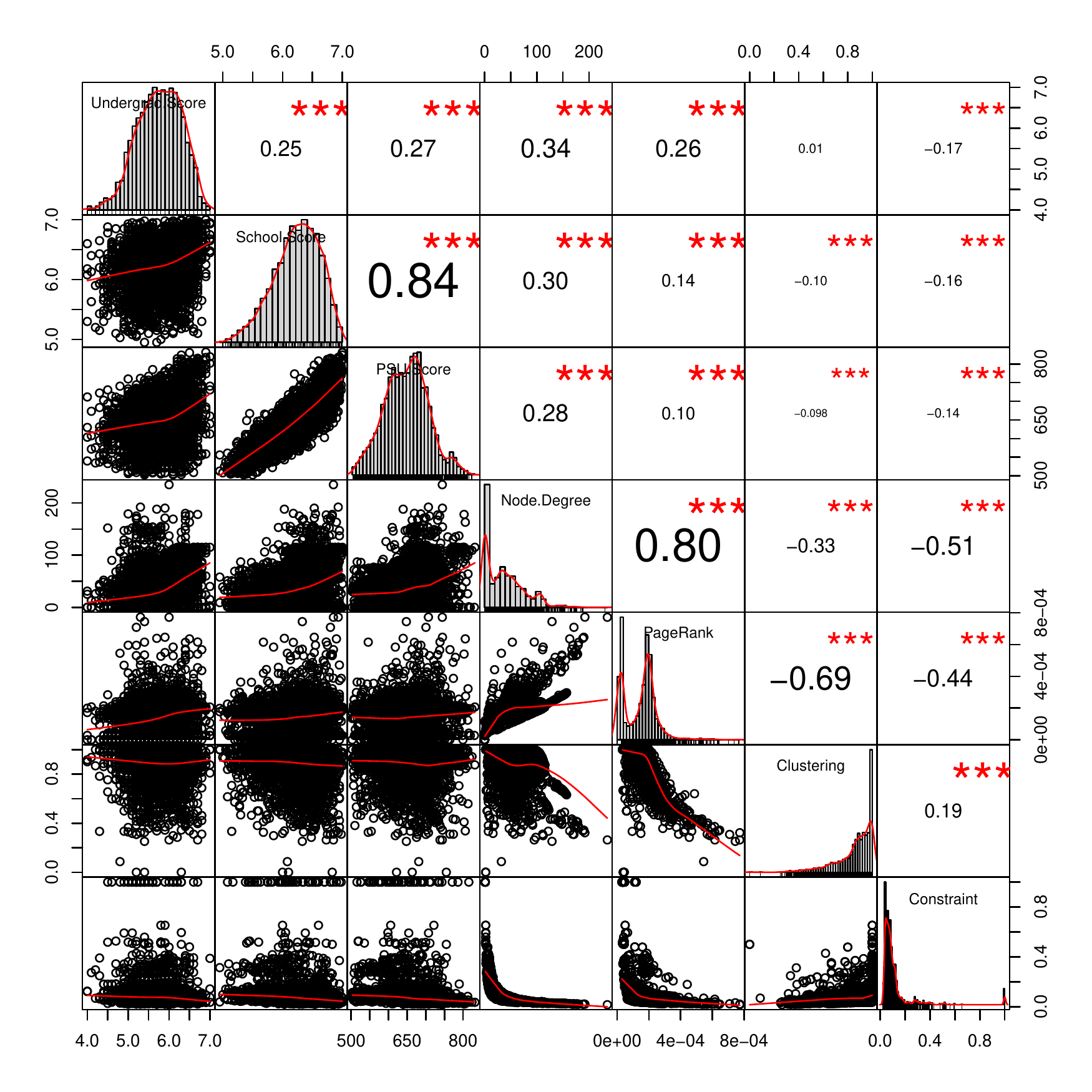}
\caption{Correlation plot for the average GPA obtained by first-year students at the end of the first semester of 2020 and different network centralities. The diagonal shows the variable histogram, the top triangle shows the Pearson correlation and their significance, and the bottom triangle shows the scatter plot between two variables.}
\label{SM_centrality}
\end{figure} 

\clearpage
\pagebreak

\begin{figure}[h!]
\centering
\includegraphics[width=0.9\linewidth]{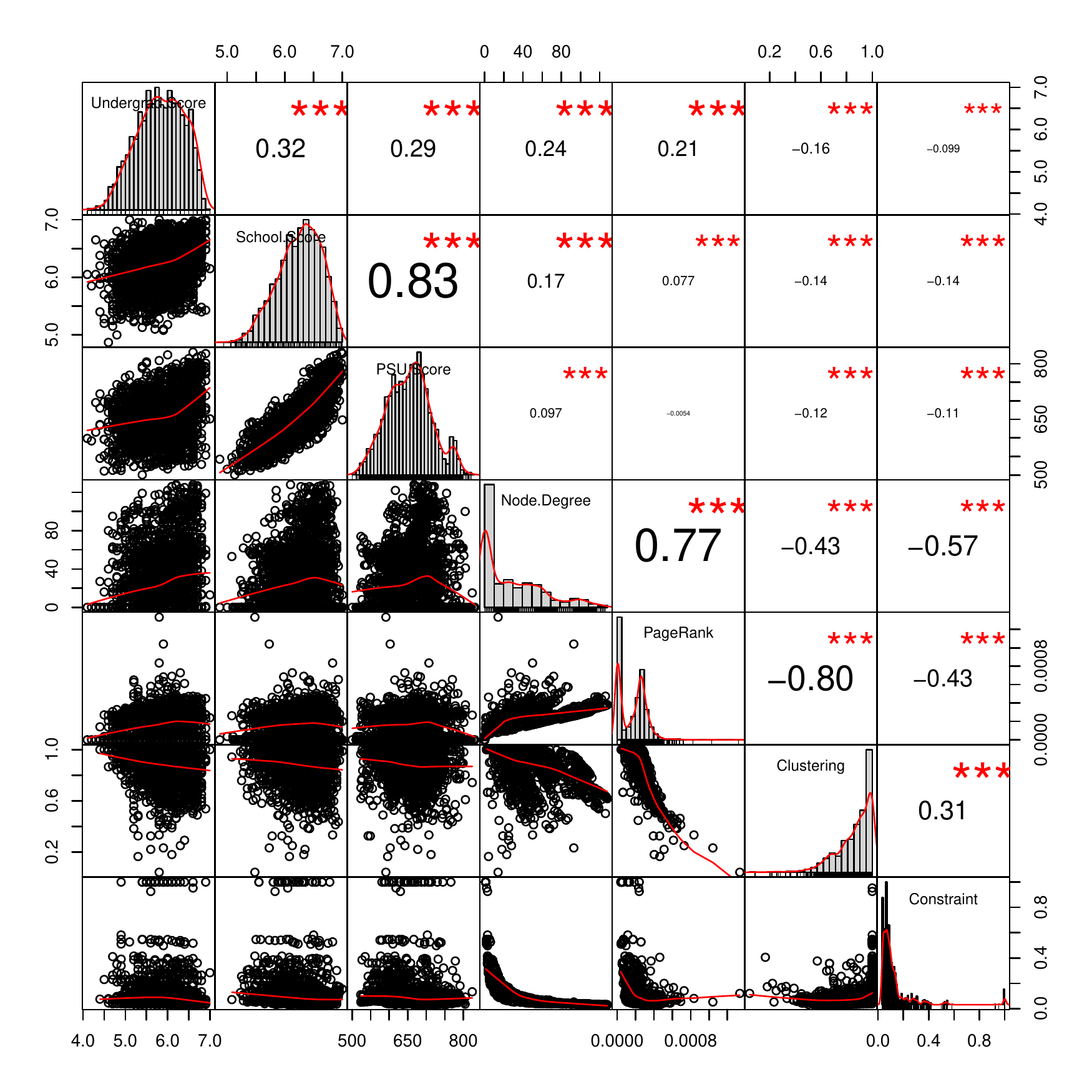}
\caption{Correlation plot for the average GPA obtained by third-year students (enrolled in 2018) at the end of the first semester of 2020 and different network centralities. The diagonal shows the variable histogram, the top triangle shows the Pearson correlation and their significance, and the bottom triangle shows the scatter plot between two variables.}
\label{SM_centrality2}
\end{figure} 

\clearpage
\pagebreak

\begin{figure}[h!]
\centering
\includegraphics[width=0.9\linewidth]{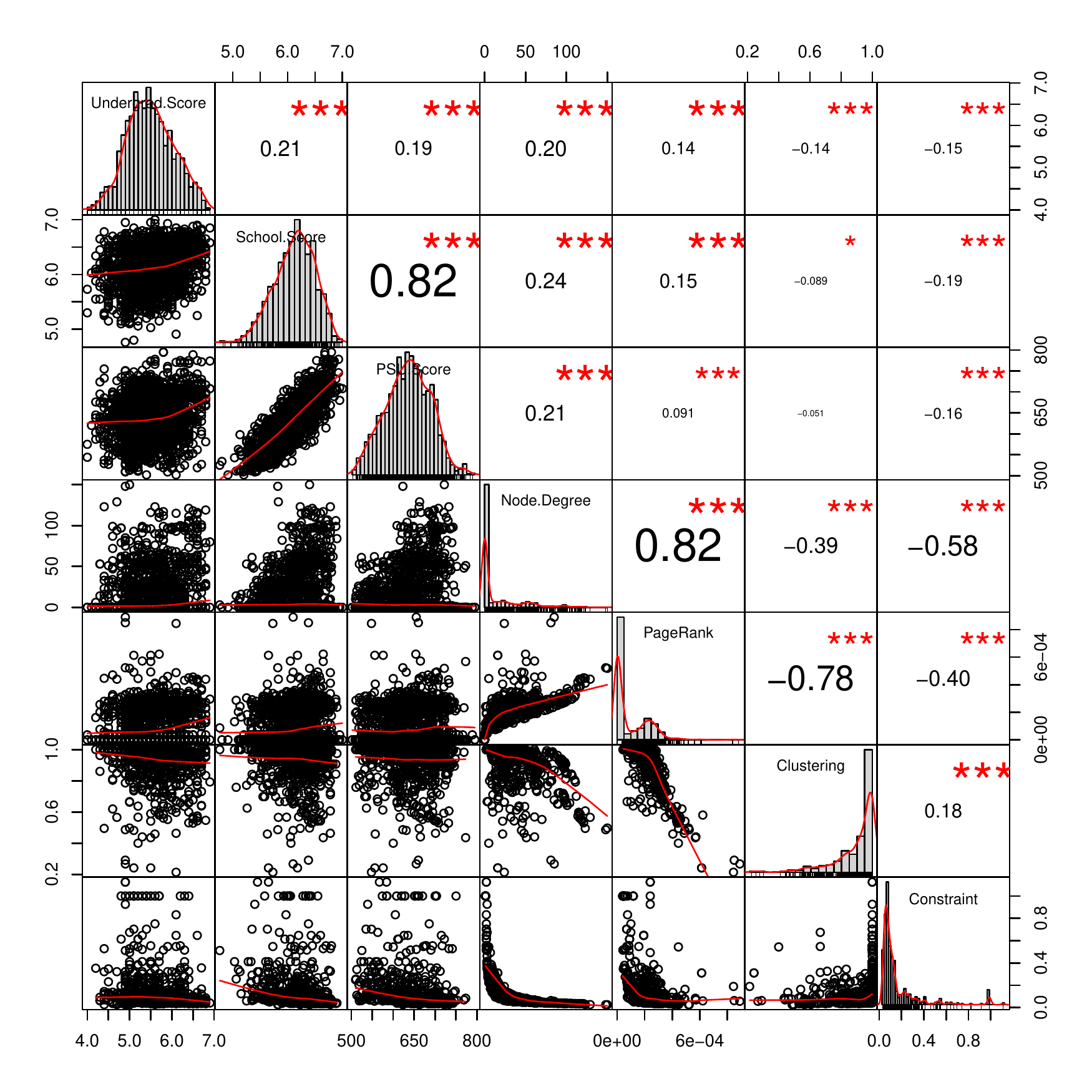}
\caption{Correlation plot for the average GPA obtained by third-year students (enrolled before 2018) at the end of the first semester of 2020 and different network centralities. The diagonal shows the variable histogram, the top triangle shows the Pearson correlation and their significance, and the bottom triangle shows the scatter plot between two variables.}
\label{SM_centrality3}
\end{figure} 

\clearpage
\pagebreak

\section{Supplementary results and robustness checks}

\subsection{Degree centrality captures forum and individual's features}

\begin{figure}[h!]
\centering
\includegraphics[width=0.99\linewidth]{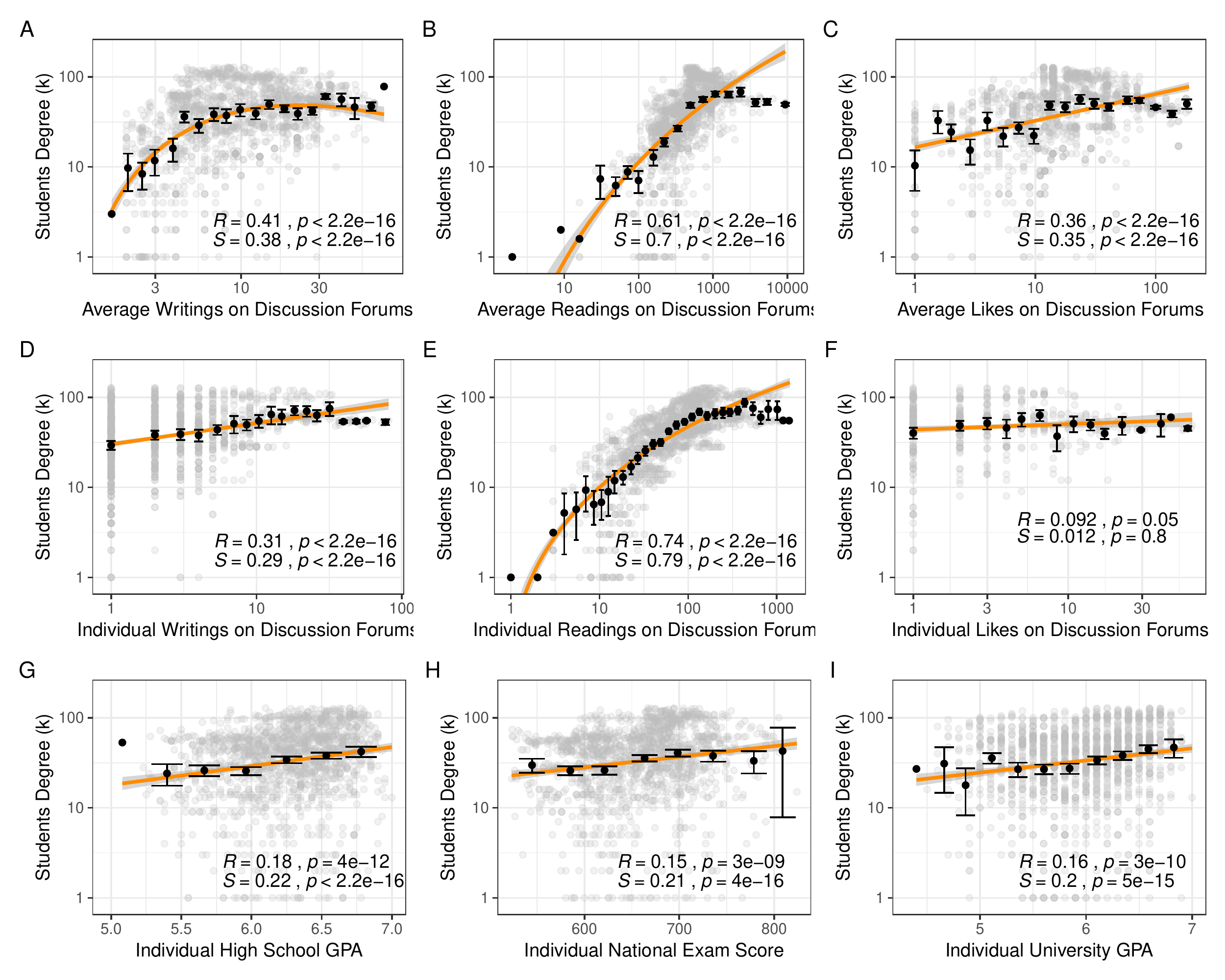}
\caption{Node degree captures forums and individual's features for third-year students enrolled in 2018. $R$ and $S$ represent Pearson and Spearman correlation respectively and $p$ is their associated p-value. For the regression models see Table \ref{degree_SM1}. Note that models do not exhibit a considerable variance inflation factor (Table \ref{vif_degree_2})}
\label{degree_SM1}
\end{figure}

\clearpage
\pagebreak

\begin{figure}[h!]
\centering
\includegraphics[width=0.99\linewidth]{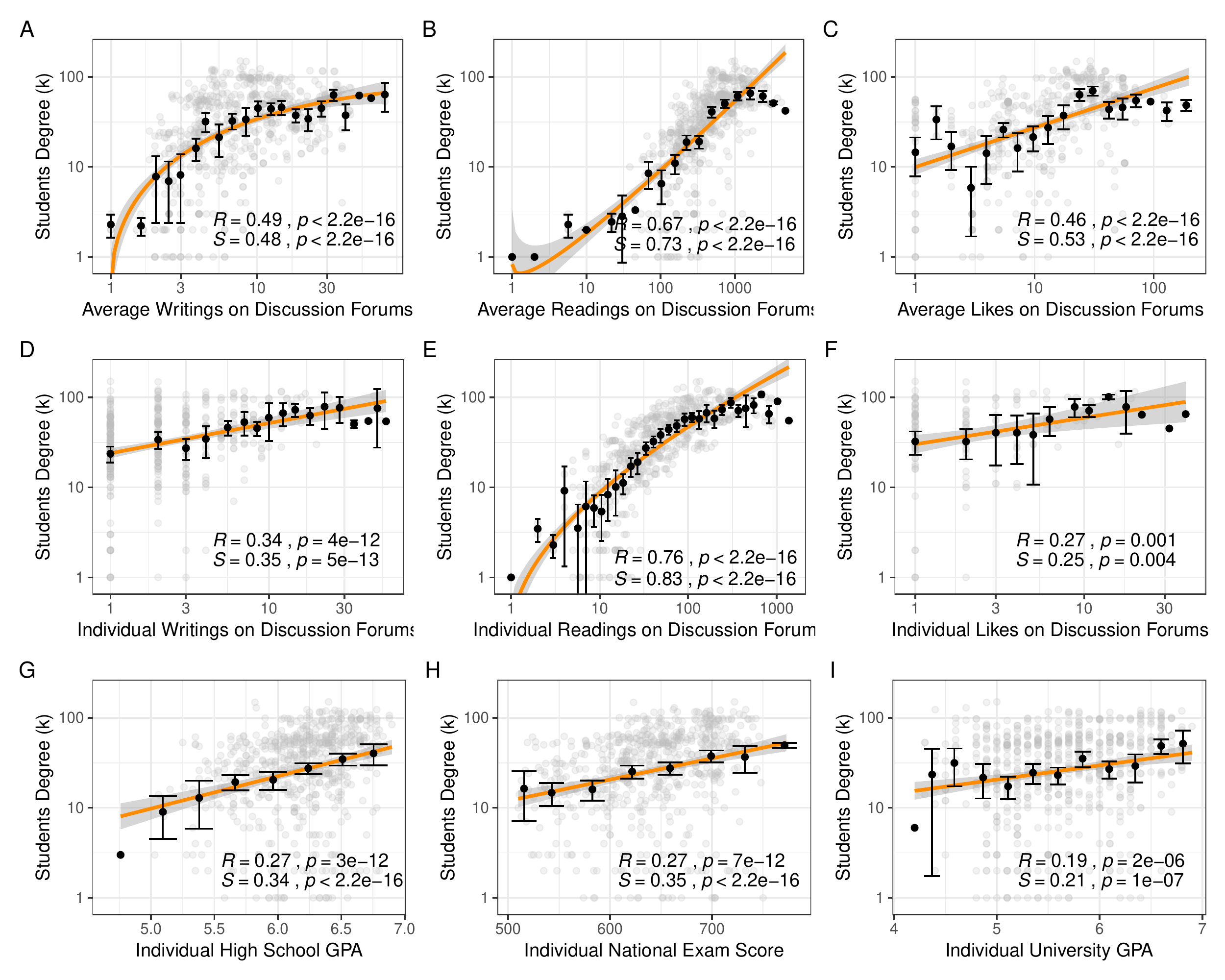}
\caption{Node degree captures forums and individual's features for third-year students enrolled before 2018. $R$ and $S$ represent Pearson and Spearman correlation respectively and $p$ is their associated p-value. For the regression models see Table \ref{degree_table}. Note that models do not exhibit a considerable variance inflation factor (Table \ref{vif_degree_3})}
\label{degree_SM2}
\end{figure}

\clearpage
\pagebreak

\begin{landscape}

\begin{table}[!htbp] \centering 
  \caption{Linear regression models for node degree.} 
  \label{degree_table} 
\tiny 
\begin{tabular}{@{\extracolsep{5pt}}lcccccc} 
\\[-1.8ex]\hline 
\hline \\[-1.8ex] 
 & \multicolumn{6}{c}{\textit{Dependent variable:}} \\ 
\cline{2-7} 
\\[-1.8ex] & \multicolumn{6}{c}{Node degree} \\ 
 & \multicolumn{2}{c}{Group 1} & \multicolumn{2}{c}{Group 2} & \multicolumn{2}{c}{Group 3} \\ 
\\[-1.8ex] & (1) & (2) & (3) & (4) & (5) & (6)\\ 
\hline \\[-1.8ex] 
 Forum's Ave. Writings (z-score) & 6.206$^{***}$ & 9.900$^{***}$ & 4.034$^{***}$ & $-$0.726 & $-$0.188 & 8.975$^{***}$ \\ 
  & (0.709) & (0.946) & (1.115) & (1.578) & (1.276) & (1.833) \\ 
  & & & & & & \\ 
 Squared Forum's Ave. Writings (z-score) & $-$2.701$^{***}$ & $-$1.976$^{***}$ & $-$1.999$^{***}$ & $-$0.158 & $-$0.266 & $-$0.956$^{***}$ \\ 
  & (0.206) & (0.207) & (0.347) & (0.359) & (0.231) & (0.264) \\ 
  & & & & & & \\ 
 Forum's Ave. Readings (z-score) & 8.634$^{***}$ & 6.175$^{***}$ & 18.793$^{***}$ & 8.423$^{***}$ & 17.534$^{***}$ & 7.675$^{***}$ \\ 
  & (0.745) & (0.956) & (1.364) & (1.752) & (1.319) & (1.920) \\ 
  & & & & & & \\ 
 Squared Forum's Ave. Readings (z-score) & $-$0.625$^{***}$ & $-$0.715$^{***}$ & $-$1.441$^{***}$ & $-$0.425$^{***}$ & $-$1.842$^{***}$ & $-$0.776$^{***}$ \\ 
  & (0.185) & (0.191) & (0.125) & (0.133) & (0.199) & (0.232) \\ 
  & & & & & & \\ 
 Forum's Ave. Likes (z-score) & 8.793$^{***}$ & 3.692$^{***}$ & $-$0.924$^{*}$ & 1.014$^{*}$ & 1.782$^{***}$ & 0.885 \\ 
  & (0.454) & (0.670) & (0.543) & (0.519) & (0.538) & (0.591) \\ 
  & & & & & & \\ 
 Student's Ave. Writings (z-score) & $-$6.187$^{***}$ & 0.077 & $-$7.765$^{***}$ & 1.035 & $-$4.175$^{***}$ & 0.178 \\ 
  & (0.534) & (0.523) & (0.803) & (0.656) & (0.841) & (0.864) \\ 
  & & & & & & \\ 
 Student's Ave. Readings (z-score) & 37.786$^{***}$ & 42.958$^{***}$ & 35.859$^{***}$ & 32.964$^{***}$ & 25.387$^{***}$ & 24.933$^{***}$ \\ 
  & (0.907) & (0.963) & (1.187) & (1.047) & (1.197) & (1.342) \\ 
  & & & & & & \\ 
 Squared Student's Ave. Readings (z-score) & $-$4.032$^{***}$ & $-$4.101$^{***}$ & $-$3.897$^{***}$ & $-$3.422$^{***}$ & $-$1.953$^{***}$ & $-$1.789$^{***}$ \\ 
  & (0.124) & (0.119) & (0.171) & (0.140) & (0.127) & (0.134) \\ 
  & & & & & & \\ 
 Student's Ave. Likes (z-score) & 0.297 & $-$0.105 & $-$0.412 & $-$0.407 & 0.211 & $-$0.760 \\ 
  & (0.376) & (0.322) & (0.481) & (0.374) & (0.476) & (0.469) \\ 
  & & & & & & \\ 
 High School GPA (z-score) & 4.014$^{***}$ & 0.201 & 5.261$^{***}$ & 0.393 & 2.820$^{***}$ & 0.241 \\ 
  & (0.411) & (0.473) & (0.540) & (0.481) & (0.621) & (0.789) \\ 
  & & & & & & \\ 
 Sex [M] & $-$0.287 & $-$0.151 & $-$0.506 & $-$0.315 & $-$1.236 & $-$1.053 \\ 
  & (0.789) & (0.708) & (1.004) & (0.736) & (1.173) & (1.219) \\ 
  & & & & & & \\ 
 N. Credits (z-score) & 0.493 & 0.353 & $-$1.846$^{***}$ & 0.562 & $-$0.977 & 0.865 \\ 
  & (0.433) & (0.447) & (0.492) & (0.478) & (0.609) & (0.770) \\ 
  & & & & & & \\ 
 Age (z-score) & 0.076 & $-$0.128 & 0.144 & 0.236 & 0.445 & $-$0.273 \\ 
  & (0.379) & (0.330) & (0.518) & (0.359) & (0.623) & (0.611) \\ 
  & & & & & & \\ 
 Constant & 45.591$^{***}$ & 54.080$^{***}$ & 38.481$^{***}$ & 44.005$^{***}$ & 20.374$^{***}$ & 23.672$^{**}$ \\ 
  & (0.565) & (6.367) & (0.669) & (5.622) & (0.835) & (9.860) \\ 
  & & & & & & \\ 
\hline \\[-1.8ex] 
Degree program f. e. & No & Yes & No & Yes & No & Yes \\ 
High school f.e. & No & Yes & No & Yes & No & Yes \\ 
Family income quintil f.e. & No & Yes & No & Yes & No & Yes \\ 
Preference ranking & No & Yes & No & Yes & No & Yes \\ 
Degree program length & No & Yes & No & Yes & No & Yes \\ 
Observations & 3,249 & 3,249 & 1,697 & 1,697 & 844 & 844 \\ 
R$^{2}$ & 0.685 & 0.805 & 0.635 & 0.872 & 0.732 & 0.847 \\ 
Adjusted R$^{2}$ & 0.684 & 0.788 & 0.632 & 0.852 & 0.727 & 0.805 \\ 
Residual Std. Error & 20.729 (df = 3235) & 16.992 (df = 2984) & 19.336 (df = 1683) & 12.253 (df = 1473) & 16.302 (df = 830) & 13.791 (df = 661) \\ 
F Statistic & 541.413$^{***}$ (df = 13; 3235) & 46.615$^{***}$ (df = 264; 2984) & 225.289$^{***}$ (df = 13; 1683) & 44.897$^{***}$ (df = 223; 1473) & 174.073$^{***}$ (df = 13; 830) & 20.113$^{***}$ (df = 182; 661) \\ 
\hline 
\hline \\[-1.8ex] 
\textit{Note:}  & \multicolumn{6}{r}{$^{*}$p$<$0.1; $^{**}$p$<$0.05; $^{***}$p$<$0.01} \\ 
\end{tabular} 
\end{table} 

\end{landscape}

\begin{table}[!htbp] \centering 
  \caption{Variance inflation factor for Group 1's node degree model} 
  \label{vif_degree_1} 
\begin{tabular}{@{\extracolsep{5pt}} cccc} 
\\[-1.8ex]\hline 
\hline \\[-1.8ex] 
 & GVIF & Df & GVIF$\hat{\mkern6mu}$(1/(2\textasteriskcentered Df)) \\ 
\hline \\[-1.8ex] 
scale(mean\_add\_entry) & $9.836$ & $1$ & $3.136$ \\ 
I(scale(mean\_add\_entry)$\hat{\mkern6mu}$2) & $3.421$ & $1$ & $1.850$ \\ 
scale(mean\_read\_entry) & $10.098$ & $1$ & $3.178$ \\ 
I(scale(mean\_read\_entry)$\hat{\mkern6mu}$2) & $4.131$ & $1$ & $2.033$ \\ 
scale(mean\_rate\_entry) & $5.432$ & $1$ & $2.331$ \\ 
scale(n\_write\_f) & $3.282$ & $1$ & $1.812$ \\ 
scale(mark\_entry\_read) & $11.094$ & $1$ & $3.331$ \\ 
I(scale(mark\_entry\_read)$\hat{\mkern6mu}$2) & $5.362$ & $1$ & $2.316$ \\ 
scale(rate\_entry) & $1.284$ & $1$ & $1.133$ \\ 
scale(Prom..Notas) & $2.544$ & $1$ & $1.595$ \\ 
Género & $1.405$ & $1$ & $1.185$ \\ 
scale(sum\_Créditos) & $2.110$ & $1$ & $1.453$ \\ 
scale(Age) & $1.224$ & $1$ & $1.106$ \\ 
Quintil & $1.357$ & $3$ & $1.052$ \\ 
Preferencia & $2.407$ & $9$ & $1.050$ \\ 
Colegio.Comuna & $338.664$ & $172$ & $1.017$ \\ 
Carrera & $151,524.200$ & $67$ & $1.093$ \\ 
\hline \\[-1.8ex] 
\end{tabular} 
\end{table} 

\begin{table}[!htbp] \centering 
  \caption{Variance inflation factor for Group 2's node degree model} 
  \label{vif_degree_2} 
\begin{tabular}{@{\extracolsep{5pt}} cccc} 
\\[-1.8ex]\hline 
\hline \\[-1.8ex] 
 & GVIF & Df & GVIF$\hat{\mkern6mu}$(1/(2\textasteriskcentered Df)) \\ 
\hline \\[-1.8ex] 
scale(mean\_add\_entry) & $29.256$ & $1$ & $5.409$ \\ 
I(scale(mean\_add\_entry)$\hat{\mkern6mu}$2) & $15.017$ & $1$ & $3.875$ \\ 
scale(mean\_read\_entry) & $41.939$ & $1$ & $6.476$ \\ 
I(scale(mean\_read\_entry)$\hat{\mkern6mu}$2) & $22.682$ & $1$ & $4.763$ \\ 
scale(mean\_rate\_entry) & $3.681$ & $1$ & $1.918$ \\ 
scale(n\_write\_f) & $6.203$ & $1$ & $2.490$ \\ 
scale(mark\_entry\_read) & $15.208$ & $1$ & $3.900$ \\ 
I(scale(mark\_entry\_read)$\hat{\mkern6mu}$2) & $8.064$ & $1$ & $2.840$ \\ 
scale(rate\_entry) & $2.084$ & $1$ & $1.444$ \\ 
scale(Prom..Notas) & $2.326$ & $1$ & $1.525$ \\ 
Género & $1.479$ & $1$ & $1.216$ \\ 
scale(sum\_Créditos) & $2.475$ & $1$ & $1.573$ \\ 
scale(Age) & $1.291$ & $1$ & $1.136$ \\ 
Quintil & $1.596$ & $3$ & $1.081$ \\ 
Preferencia & $6.645$ & $8$ & $1.126$ \\ 
Colegio.Comuna & $8,718.344$ & $140$ & $1.033$ \\ 
Carrera & $3,523,250.000$ & $59$ & $1.136$ \\ 
\hline \\[-1.8ex] 
\end{tabular} 
\end{table} 

\begin{table}[!htbp] \centering 
  \caption{Variance inflation factor for Group 3's node degree model} 
  \label{vif_degree_3} 
\begin{tabular}{@{\extracolsep{5pt}} cccc} 
\\[-1.8ex]\hline 
\hline \\[-1.8ex] 
 & GVIF & Df & GVIF$\hat{\mkern6mu}$(1/(2\textasteriskcentered Df)) \\ 
\hline \\[-1.8ex] 
scale(mean\_add\_entry) & $20.314$ & $1$ & $4.507$ \\ 
I(scale(mean\_add\_entry)$\hat{\mkern6mu}$2) & $8.578$ & $1$ & $2.929$ \\ 
scale(mean\_read\_entry) & $23.055$ & $1$ & $4.802$ \\ 
I(scale(mean\_read\_entry)$\hat{\mkern6mu}$2) & $9.705$ & $1$ & $3.115$ \\ 
scale(mean\_rate\_entry) & $2.560$ & $1$ & $1.600$ \\ 
scale(n\_write\_f) & $5.913$ & $1$ & $2.432$ \\ 
scale(mark\_entry\_read) & $13.661$ & $1$ & $3.696$ \\ 
I(scale(mark\_entry\_read)$\hat{\mkern6mu}$2) & $7.672$ & $1$ & $2.770$ \\ 
scale(rate\_entry) & $1.851$ & $1$ & $1.360$ \\ 
scale(Prom..Notas) & $2.705$ & $1$ & $1.645$ \\ 
Género & $1.628$ & $1$ & $1.276$ \\ 
scale(sum\_Créditos) & $2.375$ & $1$ & $1.541$ \\ 
scale(Age) & $1.469$ & $1$ & $1.212$ \\ 
Quintil & $4.458$ & $5$ & $1.161$ \\ 
factor(Preferencia) & $5.841$ & $8$ & $1.117$ \\ 
Colegio.Comuna & $191,814.900$ & $96$ & $1.065$ \\ 
Carrera & $7,561,474.000$ & $60$ & $1.141$ \\ 
\hline \\[-1.8ex] 
\end{tabular} 
\end{table}

\clearpage
\pagebreak

\subsection{Main model table}

\begin{table}[!htbp] \centering 
  \caption{Main hierarchical regression models represented in main Figure \ref{fig1M}.} 
  \label{main_table} 
    \scriptsize
\begin{tabular}{@{\extracolsep{5pt}}lccc} 
\\[-1.8ex]\hline 
\hline \\[-1.8ex] 
 & \multicolumn{3}{c}{\textit{Dependent variable:}} \\ 
\cline{2-4} 
\\[-1.8ex] & \multicolumn{3}{c}{Average GPA 1-2020} \\ 
   & First-year students & Third-year students & Third-year students\\ 
  & (enrolled in 2020) & (enrolled in 2018) & (enrolled before 2018) \\ 
\\[-1.8ex] & (1) & (2) & (3)\\ 
\hline \\[-1.8ex] 
 Sex [M] & $-$0.005 & $-$0.117$^{***}$ & $-$0.103$^{***}$ \\ 
  & (0.015) & (0.017) & (0.023) \\ 
  & & & \\ 
 N. Credits (z-score) & $-$0.049$^{***}$ & 0.029$^{***}$ & 0.068$^{***}$ \\ 
  & (0.009) & (0.011) & (0.012) \\ 
  & & & \\ 
 Forum Writing (z-score) & 0.094$^{***}$ & 0.075$^{***}$ & 0.035 \\ 
  & (0.011) & (0.017) & (0.023) \\ 
  & & & \\ 
 Forum Reading (z-score) & 0.026$^{*}$ & 0.020 & 0.072$^{**}$ \\ 
  & (0.014) & (0.021) & (0.028) \\ 
  & & & \\ 
 Node Degree (z-score) & 0.081$^{***}$ & 0.133$^{***}$ & $-$0.018 \\ 
  & (0.014) & (0.024) & (0.029) \\ 
  & & & \\ 
 High School GPA (z-score) & 0.102$^{***}$ & 0.119$^{***}$ & 0.081$^{***}$ \\ 
  & (0.009) & (0.011) & (0.015) \\ 
  & & & \\ 
 Age (z-score) & 0.004 & 0.005 & $-$0.023$^{**}$ \\ 
  & (0.007) & (0.008) & (0.011) \\ 
  & & & \\ 
 Node Degree (z-score) x High School GPA (z-score) & $-$0.034$^{***}$ & $-$0.024$^{**}$ & 0.005 \\ 
  & (0.009) & (0.011) & (0.015) \\ 
  & & & \\ 
 Constant & 5.787$^{***}$ & 5.868$^{***}$ & 5.579$^{***}$ \\ 
  & (0.037) & (0.043) & (0.044) \\ 
  & & & \\ 
\hline \\[-1.8ex] 
Random slope for node degree in each degree program & Yes & Yes & Yes \\ 
Degree program random intercept & Yes & Yes & Yes \\ 
High school random intercept & Yes & Yes & Yes \\ 
Commune random intercept & Yes & Yes & Yes \\ 
\hline
Observations & 3,585 & 2,296 & 1,647 \\ 
Log Likelihood & $-$1,771.267 & $-$990.039 & $-$947.311 \\ 
Akaike Inf. Crit. & 3,572.534 & 2,010.079 & 1,924.622 \\ 
Bayesian Inf. Crit. & 3,665.302 & 2,096.163 & 2,005.722 \\ 
\hline 
\hline \\[-1.8ex] 
\textit{Note:}  & \multicolumn{3}{r}{$^{*}$p$<$0.1; $^{**}$p$<$0.05; $^{***}$p$<$0.01} \\ 
\end{tabular} 
\end{table} 

\clearpage
\pagebreak

\subsection{Main model random slopes}

\begin{figure}[h!]
\centering
\includegraphics[width=0.99\linewidth]{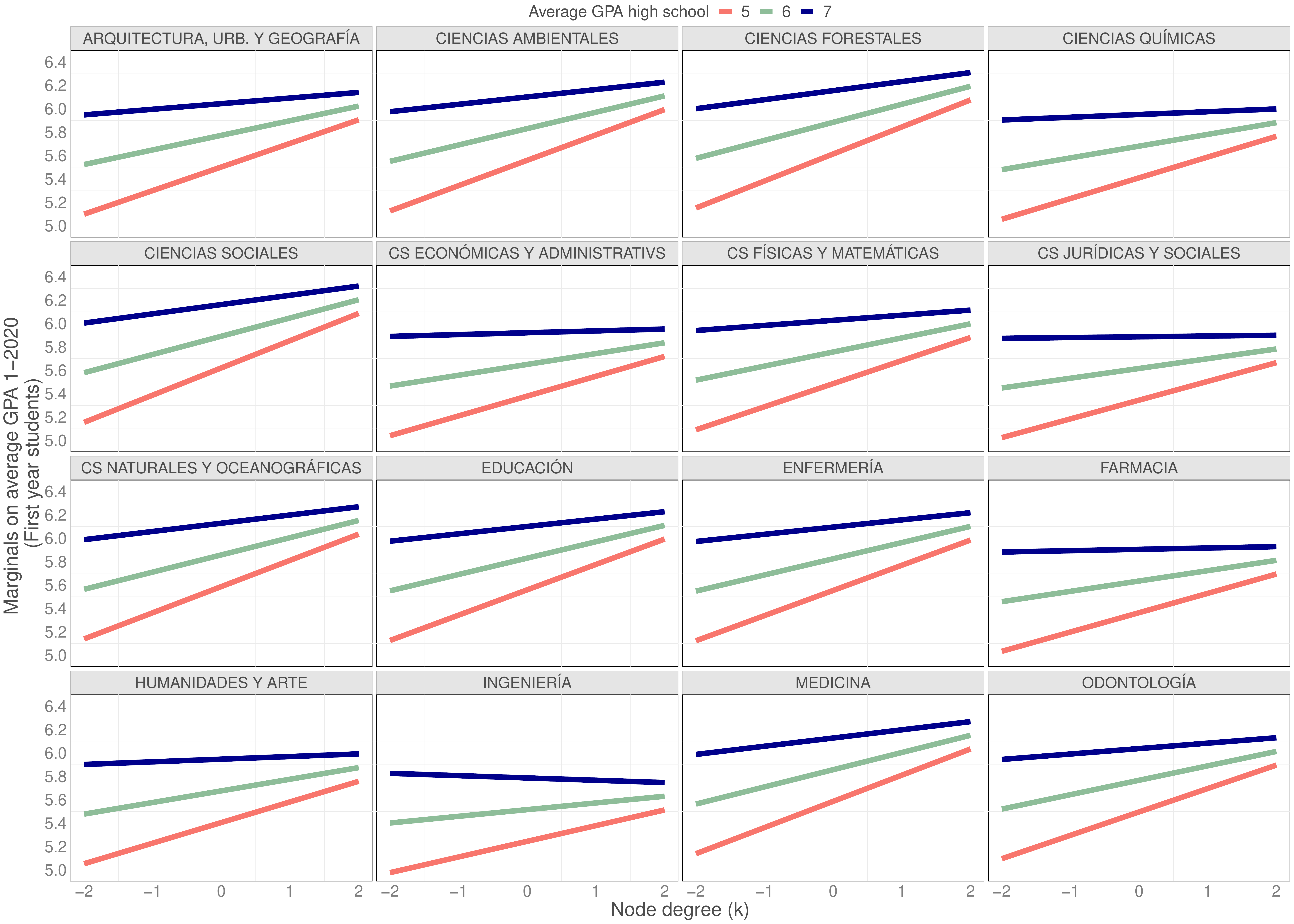}
\caption{Hierarchical regression node degree random slope for different knowledge areas. The dependent variable is the average GPA obtained by first-year students at the end of the first semester of 2020. Colors represent different high school GPAs.}
\label{fig4M}
\end{figure} 

\clearpage
\pagebreak

\subsection{Main model non-standardized coefficients}

\begin{figure}[h!]
\centering
\includegraphics[width=0.99\linewidth]{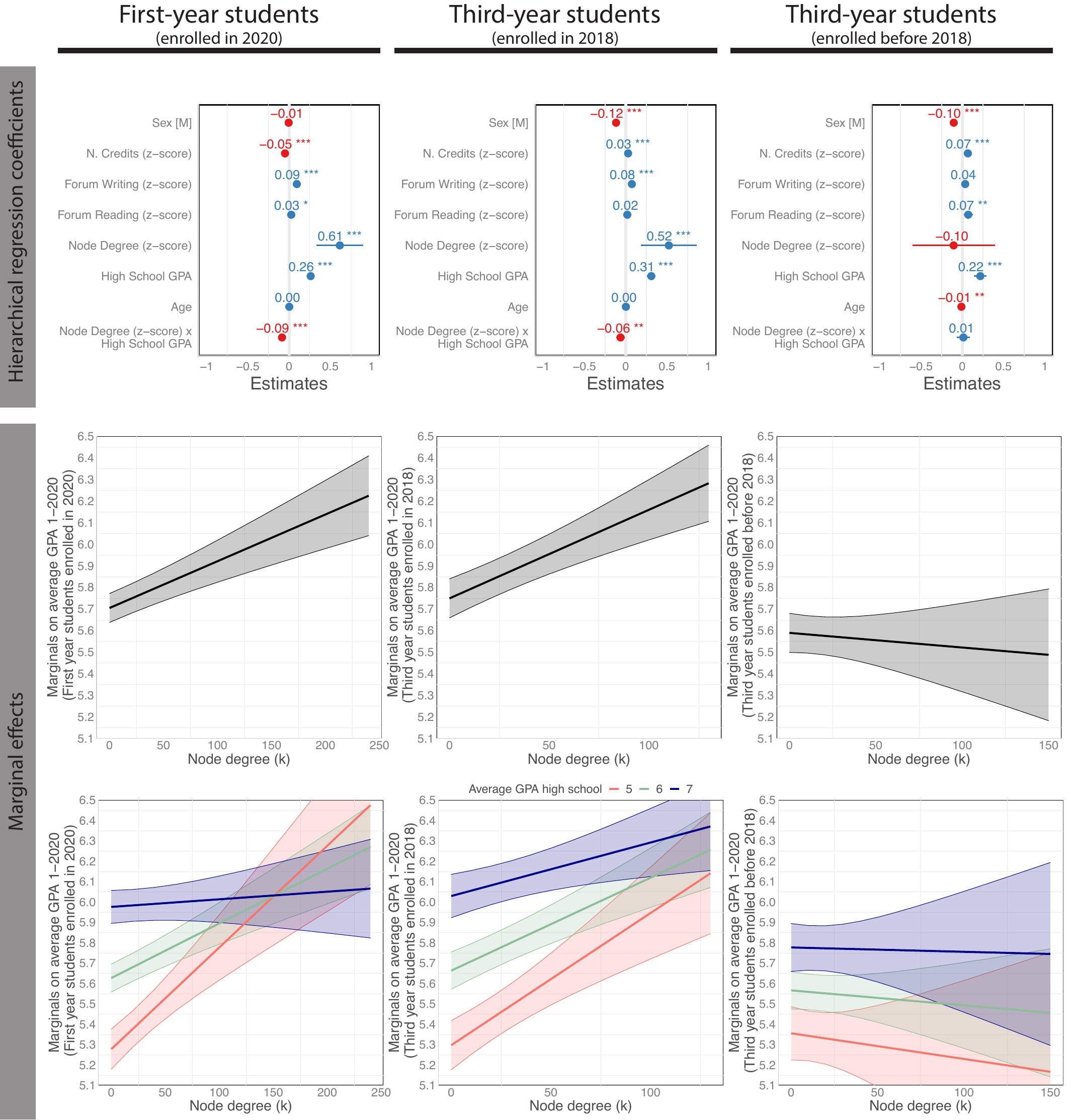}
\caption{Hierarchical regression models for the average GPA obtained by three samples of students at the end of the first semester of 2020. The first column refers to first-year students enrolled in 2020.  The second column refers to third-year students enrolled in 2018, i.e., students who do not have any failed subject. The third column refers to third-year students enrolled before 2018, i.e., students taking overdue subjects. The first row shows the regression coefficients. The second row depicts the marginal effects on node degree (k), and the third row depicts the same marginal effects for different high school GPAs.}
\label{main_sm}
\end{figure}

\clearpage
\pagebreak

\subsection{Main model predicted vs actual average GPA}

\begin{figure}[h!]
\centering
\includegraphics[width=0.99\linewidth]{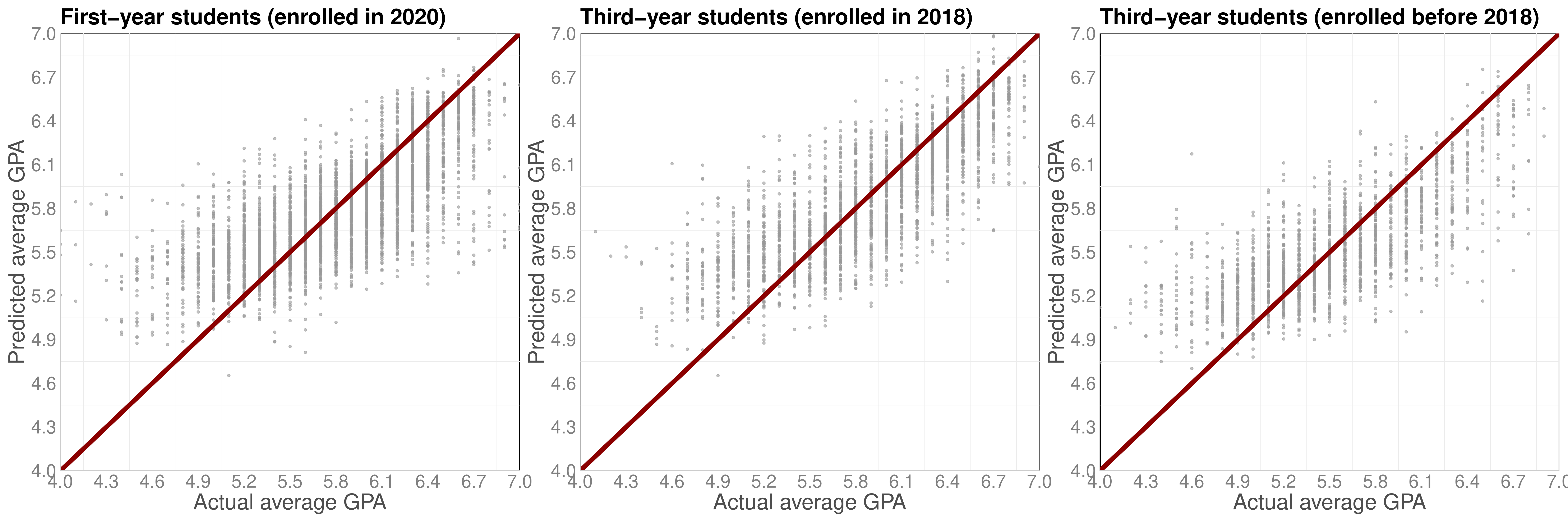}
\caption{Predicted vs actual plot for the average GPA obtained by students at the end of the first semester of 2020. A) First-year students entering university for the first time. B) Third-year students who do not have any failed subject. C) Third-year students taking overdue subjects. The red line represents the perfect fit.}
\label{SM_diag_plot}
\end{figure} 
\clearpage
\pagebreak

\subsection{The effect of node degree on undergrad GPA without the interaction term}

\begin{figure}[h!]
\centering
\includegraphics[width=0.99\linewidth]{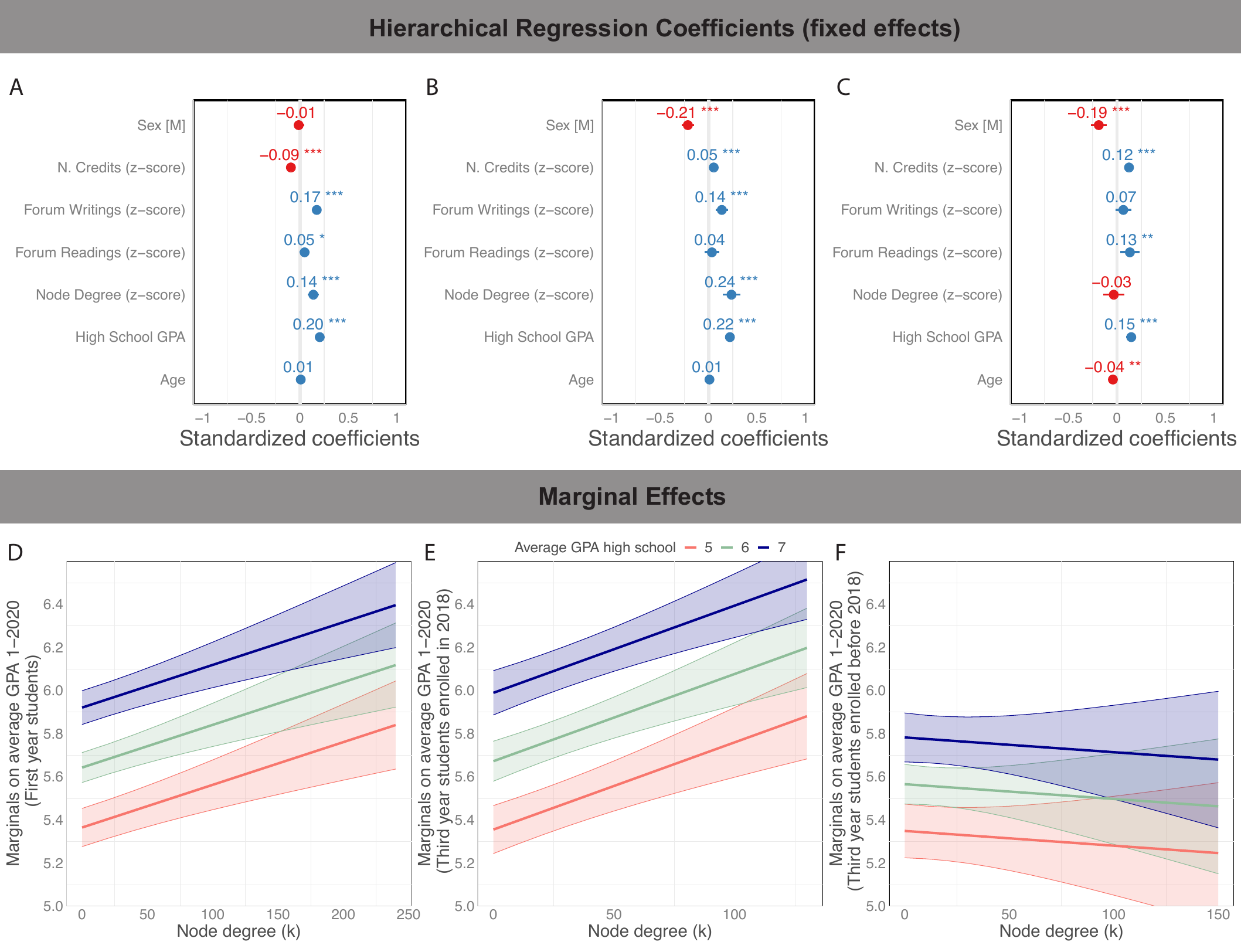}
\caption{Hierarchical regression models for the average GPA obtained by three samples of students at the end of the first semester of 2020 not including the interaction term between node degree and high school GPA. The first column refers to first-year students enrolled in 2020. The second column refers to third-year students enrolled in 2018, i.e., students who do not have any failed subject. The third column refers to third-year students enrolled before 2018, i.e., students taking overdue subjects. We show the standardized regression coefficients (first row) to compare the effects between different independent variables. The second row depicts the marginal effects on the standardized number of students' connections in the information co-exposure network built from online class forums readings, quantified as the node degree (k) for different high school GPAs (colors).}
\label{SM_no_interaction}
\end{figure}

\begin{table}[!htbp] \centering 
\scriptsize
  \caption{ANOVA testing the improvement at including the interaction term between node degree and high school GPA for each student sample (model 1, model 2, and model 3). The tested models are depicted in Fig. \ref{fig1M} (with interaction) and Fig.\ref{SM_no_interaction} (without interaction). For samples 1 and 2 the interaction term improves the models' performance.} 
  \label{anova_main} 
\begin{tabular}{@{\extracolsep{5pt}} ccccccccc} 
\\[-1.8ex]\hline 
\hline \\[-1.8ex] 
 & Df & AIC & BIC & logLik & deviance & Chisq & Chi Df & Pr(\textgreater Chisq) \\ 
\hline \\[-1.8ex] 
model 1 no-interaction & $14$ & $3,584.574$ & $3,671.157$ & -$1,778.287$ & $3,556.574$ &  &  &  \\ 
model 1 interaction & $15$ & $3,572.534$ & $3,665.302$ & -$1,771.267$ & $3,542.534$ & $14.039$ & $1$ & $0.0002$ \\ 
\hline \\[-1.8ex] 
model 2 no-interaction & $14$ & $2,013.215$ & $2,093.560$ & -$992.608$ & $1,985.215$ & $$ & $$ &  \\ 
model 2 interaction & $15$ & $2,010.079$ & $2,096.163$ & -$990.039$ & $1,980.079$ & $5.136$ & $1$ & $0.023$ \\ 
\hline \\[-1.8ex] 
model 3 no-interaction & $14$ & $1,922.736$ & $1,998.430$ & -$947.368$ & $1,894.736$ & $$ & $$ &  \\ 
model 3 interaction & $15$ & $1,924.622$ & $2,005.722$ & -$947.311$ & $1,894.622$ & $0.114$ & $1$ & $0.736$ \\ 
\hline \\[-1.8ex]
\end{tabular} 
\end{table}

\clearpage
\pagebreak

\subsection{Clustering coefficient and university GPA}

\begin{figure}[h!]
\centering
\includegraphics[width=0.99\linewidth]{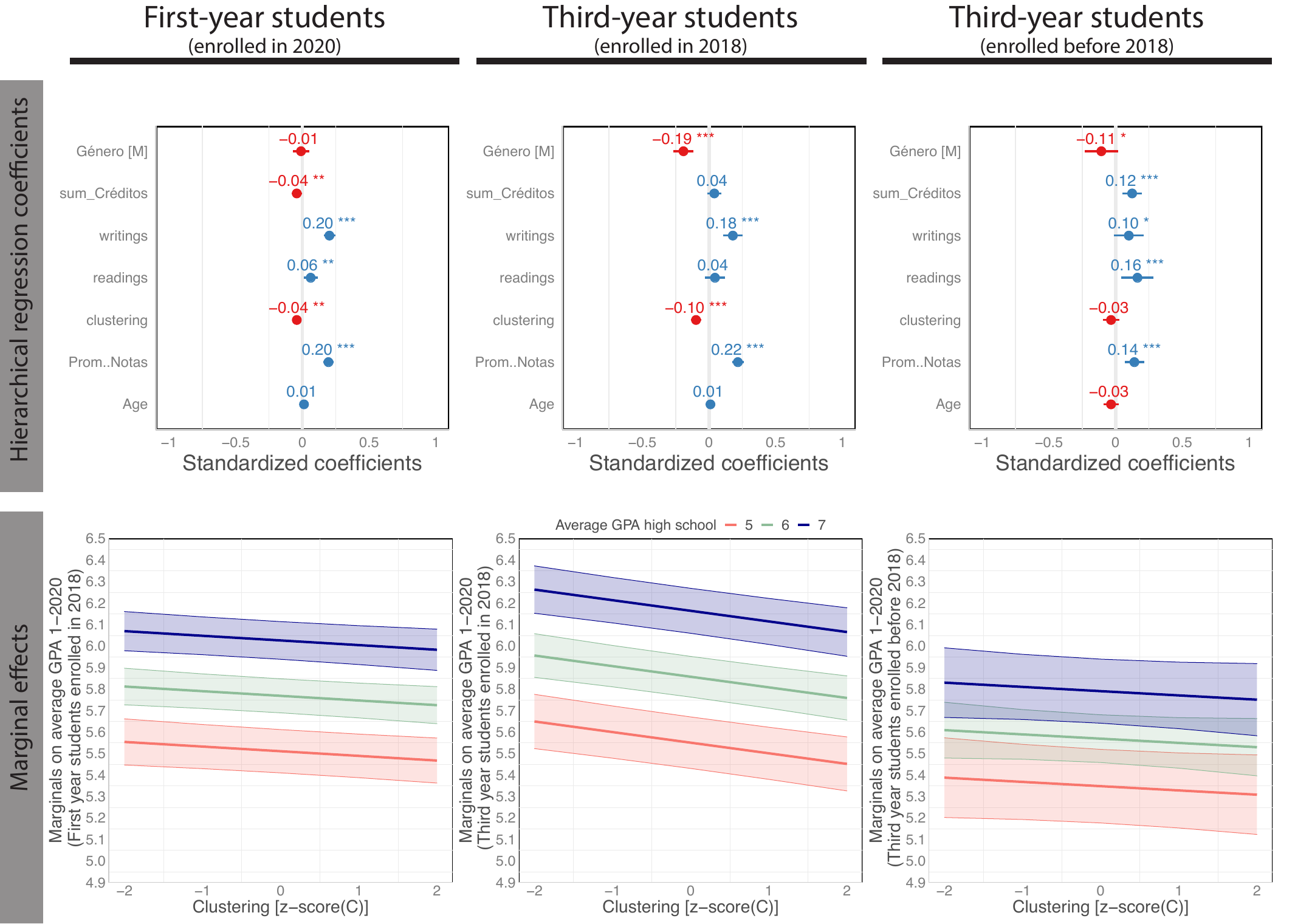}
\caption{Hierarchical regression models for the average GPA obtained by three samples of students at the end of the first semester of 2020. The first column refers to first-year students enrolled in 2020. The second column refers to third-year students enrolled in 2018, i.e., students who do not have any failed subject. The third column refers to third-year students enrolled before 2018, i.e., students taking overdue subjects. We show the standardized regression coefficients (first row) to compare the effects between different independent variables. The second row depicts the marginal effects on the standardized student's clustering coefficient (C) in the forum co-exposure network built from online class forums readings for different high school GPAs (colors), which quantifies the potential access to redundant information. (See Fig.\ref{SM_constraint} for Burt's network constraint that shows similar results).}
\label{fig3M}
\end{figure}

\clearpage
\pagebreak

\subsection{Node degree and national exam (PSU Score)}

\begin{figure}[h!]
\centering
\includegraphics[width=0.99\linewidth]{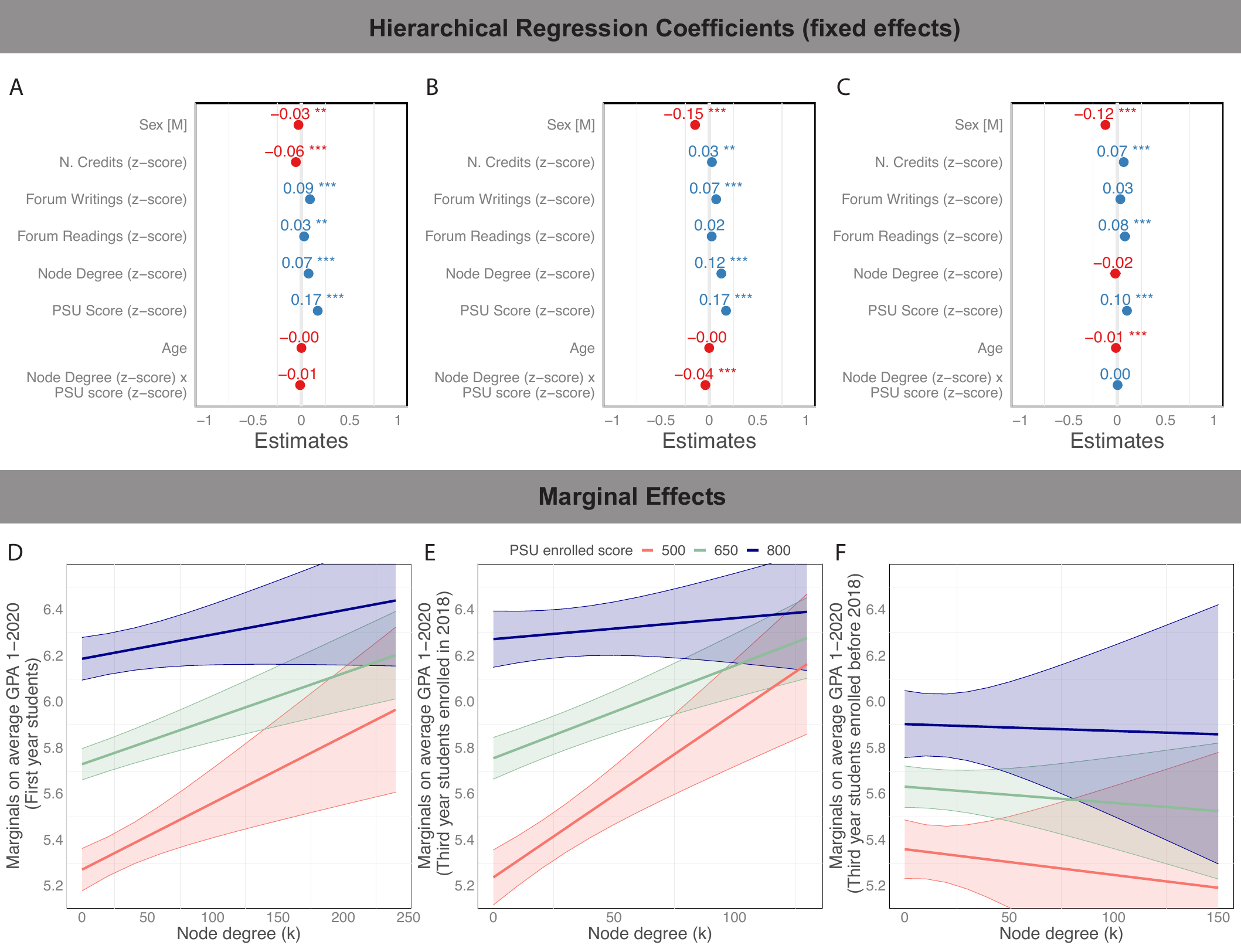}
\caption{Marginal effects for the average GPA obtained by students at the end of the first semester of 2020. A) First-year students entering university for the first time. B) Third-year students who do not have any failed subject. C) Third-year students taking overdue subjects. Colors represent high school GPA.}
\label{Res_SM3}
\end{figure}

\clearpage
\pagebreak

\subsection{Main model using PageRank}

\begin{figure}[h!]
\centering
\includegraphics[width=0.99\linewidth]{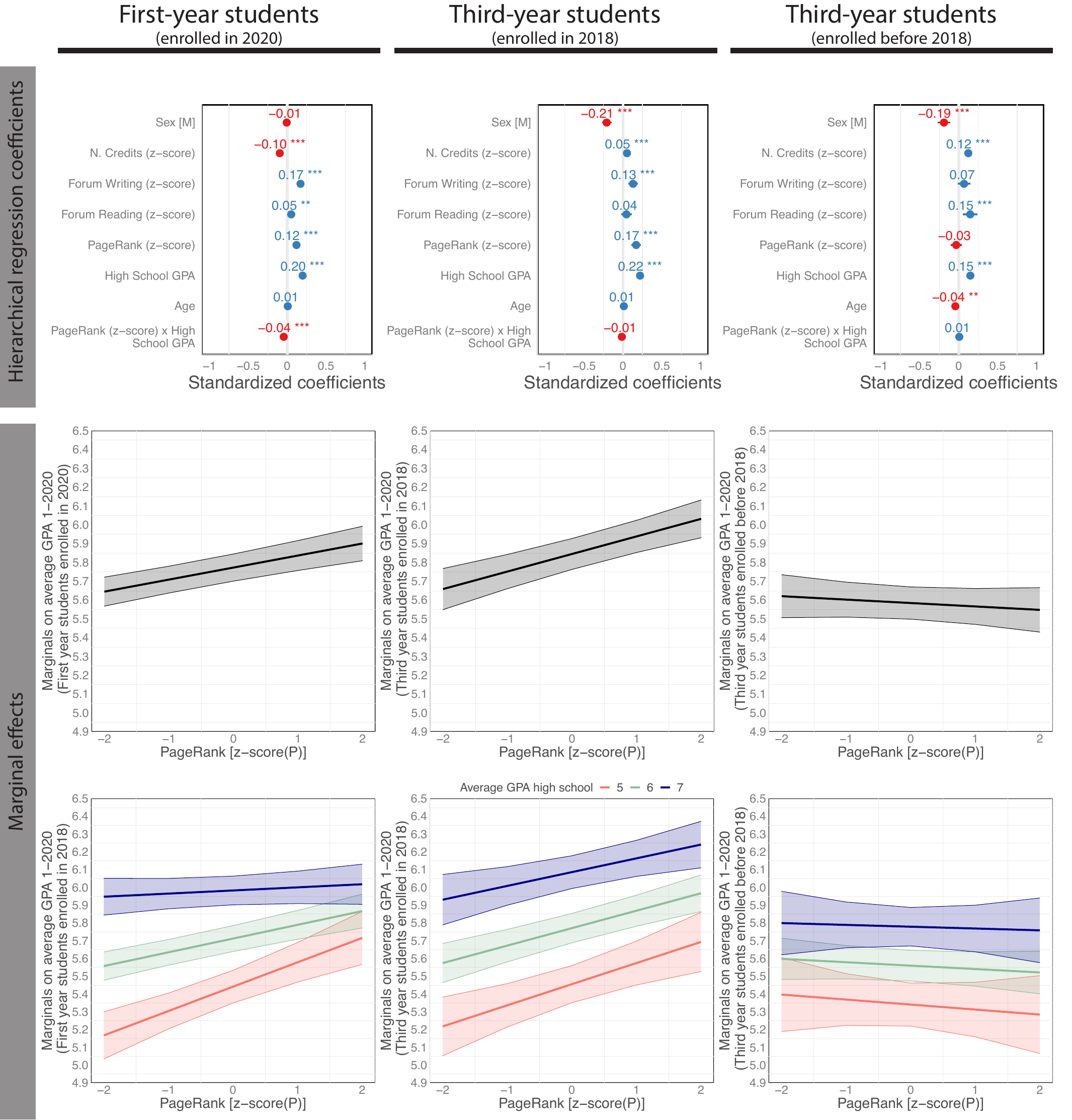}
\caption{Hierarchical regression models for the average GPA obtained by three samples of students at the end of the first semester of 2020. The first column refers to first-year students enrolled in 2020. The second column refers to third-year students enrolled in 2018, i.e., students who do not have any failed subject. The third column refers to third-year students enrolled before 2018, i.e., students taking overdue subjects. We show the standardized regression coefficients (first row) to compare the effects between different independent variables. The second row depicts the marginal effects on the standardized number of students' connections in the information co-exposure network built from online class forums readings, quantified as the PageRanl (P). The third row depicts the same marginal effects but for different high school GPAs. (See Fig.\ref{main_sm} for non-standardized regression models, Table\ref{main_table} for more details of non-standardized models, and Fig.\ref{SM_diag_plot} for the diagonal plot of each model).}
\label{SM_PageRank}
\end{figure}

\clearpage
\pagebreak

\subsection{Burt's network constraint}

\begin{figure}[h!]
\centering
\includegraphics[width=0.99\linewidth]{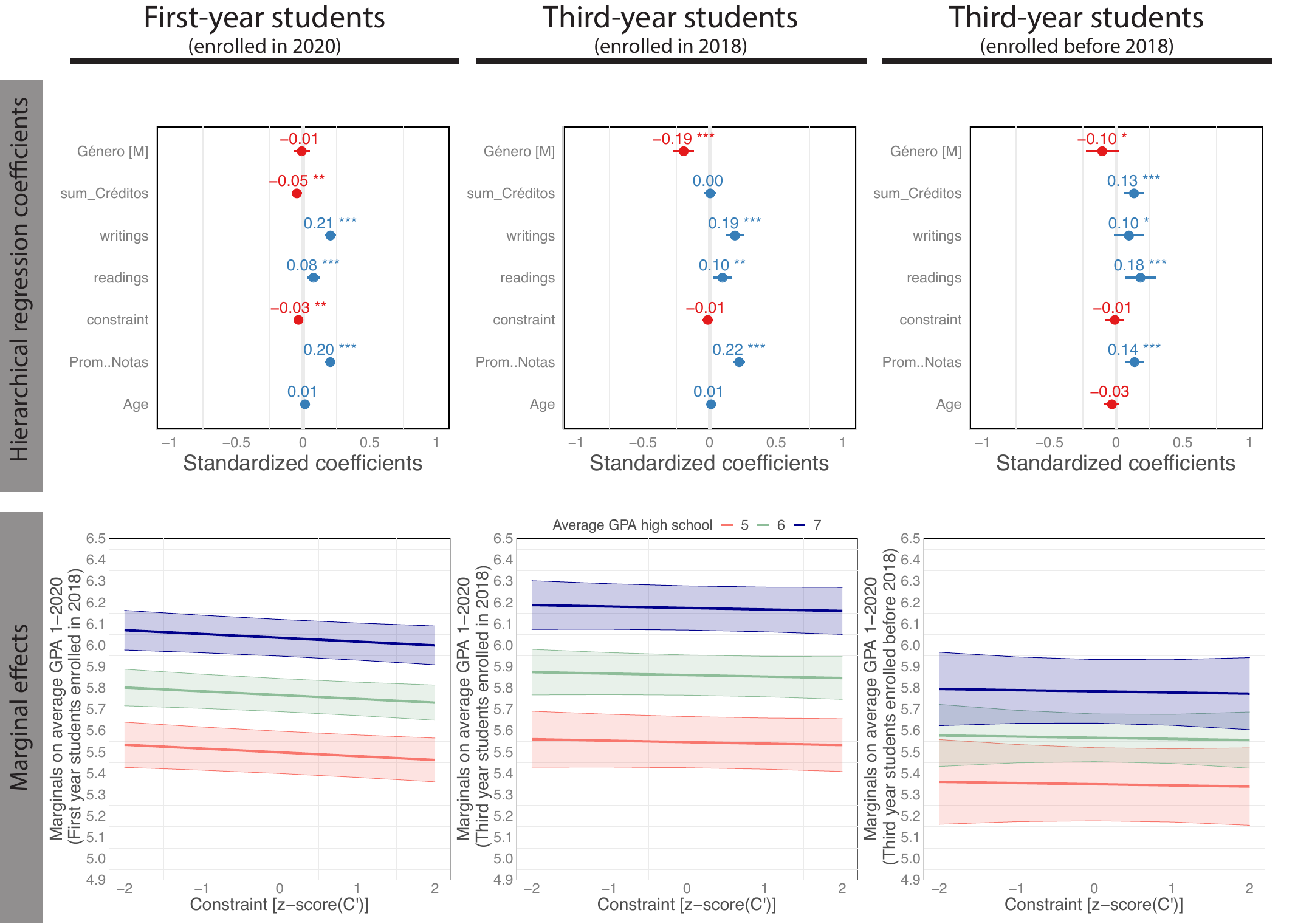}
\caption{Hierarchical regression models for the average GPA obtained by three samples of students at the end of the first semester of 2020. The first column refers to first-year students enrolled in 2020. The second column refers to third-year students enrolled in 2018, i.e., students who do not have any failed subject. The third column refers to third-year students enrolled before 2018, i.e., students taking overdue subjects. We show the standardized regression coefficients (first row) to compare the effects between different independent variables. The second row depicts the marginal effects on the standardized Burt's constraint coefficient (C') in the forum co-exposure network built from online class forums readings for different high school GPAs (colors), which quantifies the potential access to redundant information.}
\label{SM_constraint}
\end{figure}

\clearpage
\pagebreak
\subsection{Regression models with content exposure}

\begin{table}[!htbp] \centering 
  \caption{Hierarchical regression model adding content exposure for first-year students enrolled in 2020.} 
  \label{bottom_top1} 
\tiny 
\begin{tabular}{@{\extracolsep{5pt}}lcccc} 
\\[-1.8ex]\hline 
\hline \\[-1.8ex] 
 & \multicolumn{4}{c}{\textit{Dependent variable:}} \\ 
\cline{2-5} 
\\[-1.8ex] & \multicolumn{4}{c}{Average GPA 1-2020} \\ 
 & Main model & Main model and content & Main model  & Main model and content \\ 
  & &  & (no forum controls) & (no forum controls) \\ 
\\[-1.8ex] & (1) & (2) & (3) & (4)\\ 
\hline \\[-1.8ex] 
 Sex [M] & $-$0.005 & $-$0.010 & $-$0.011 & $-$0.017 \\ 
  & (0.015) & (0.015) & (0.015) & (0.015) \\ 
  & & & & \\ 
 N. Credits (z-score) & $-$0.049$^{***}$ & $-$0.064$^{***}$ & $-$0.048$^{***}$ & $-$0.061$^{***}$ \\ 
  & (0.009) & (0.010) & (0.009) & (0.010) \\ 
  & & & & \\ 
 Forum Writing (z-score) & 0.094$^{***}$ & 0.090$^{***}$ &  &  \\ 
  & (0.011) & (0.011) &  &  \\ 
  & & & & \\ 
 Forum Reading (z-score) & 0.026$^{*}$ & 0.028$^{**}$ &  &  \\ 
  & (0.014) & (0.014) &  &  \\ 
  & & & & \\ 
 Node Degree (z-score) & 0.081$^{***}$ & 0.096$^{***}$ & 0.131$^{***}$ & 0.143$^{***}$ \\ 
  & (0.014) & (0.025) & (0.014) & (0.025) \\ 
  & & & & \\ 
 Content Exposure [Low] &  & $-$0.082$^{**}$ &  & $-$0.091$^{**}$ \\ 
  &  & (0.036) &  & (0.037) \\ 
  & & & & \\ 
 Content Exposure [Medium] &  & $-$0.082$^{***}$ &  & $-$0.083$^{***}$ \\ 
  &  & (0.028) &  & (0.028) \\ 
  & & & & \\ 
 High School GPA (z-score) & 0.102$^{***}$ & 0.103$^{***}$ & 0.111$^{***}$ & 0.111$^{***}$ \\ 
  & (0.009) & (0.010) & (0.010) & (0.010) \\ 
  & & & & \\ 
 Age (z-score) & 0.004 & 0.003 & 0.006 & 0.004 \\ 
  & (0.007) & (0.007) & (0.007) & (0.007) \\ 
  & & & & \\ 
 Node Degree (z-score) x Content Exposure [Low] &  & $-$0.049$^{*}$ &  & $-$0.063$^{**}$ \\ 
  &  & (0.028) &  & (0.028) \\ 
  & & & & \\ 
 Node Degree (z-score) x Content Exposure [Medium] &  & 0.004 &  & 0.012 \\ 
  &  & (0.027) &  & (0.027) \\ 
  & & & & \\ 
 Node Degree (z-score) x High School GPA (z-score) & $-$0.034$^{***}$ & $-$0.025$^{***}$ & $-$0.030$^{***}$ & $-$0.021$^{**}$ \\ 
  & (0.009) & (0.009) & (0.009) & (0.009) \\ 
  & & & & \\ 
 Constant & 5.787$^{***}$ & 5.852$^{***}$ & 5.790$^{***}$ & 5.861$^{***}$ \\ 
  & (0.037) & (0.042) & (0.038) & (0.042) \\ 
  & & & & \\ 
\hline \\[-1.8ex] 
Random slope for node degree in each degree program & Yes & Yes & Yes & Yes \\ 
Degree program random intercept & Yes & Yes & Yes & Yes \\ 
High school random intercept & Yes & Yes & Yes & Yes \\ 
Commune random intercept & Yes & Yes & Yes & Yes \\ 
\hline
Observations & 3,585 & 3,248 & 3,585 & 3,248 \\ 
Log Likelihood & $-$1,771.267 & $-$1,580.114 & $-$1,818.845 & $-$1,619.718 \\ 
Akaike Inf. Crit. & 3,572.534 & 3,198.227 & 3,663.691 & 3,273.436 \\ 
Bayesian Inf. Crit. & 3,665.302 & 3,313.858 & 3,744.090 & 3,376.894 \\ 
\hline 
\hline \\[-1.8ex] 
\textit{Note:}  & \multicolumn{4}{r}{$^{*}$p$<$0.1; $^{**}$p$<$0.05; $^{***}$p$<$0.01} \\ 
\end{tabular} 
\end{table}

\begin{table}[!htbp] \centering 
  \caption{Hierarchical regression model adding content exposure for third-year students enrolled in 2018} 
  \label{bottom_top2} 
\tiny 
\begin{tabular}{@{\extracolsep{5pt}}lcccc} 
\\[-1.8ex]\hline 
\hline \\[-1.8ex] 
 & \multicolumn{4}{c}{\textit{Dependent variable:}} \\ 
\cline{2-5} 
\\[-1.8ex] & \multicolumn{4}{c}{Average GPA 1-2020} \\  & Main model & Main model and content & Main model  & Main model and content \\ 
  & &  & (no forum controls) & (no forum controls) \\ 
\\[-1.8ex] & (1) & (2) & (3) & (4)\\ 
\hline \\[-1.8ex] 
 Sex [M] & $-$0.117$^{***}$ & $-$0.091$^{***}$ & $-$0.122$^{***}$ & $-$0.098$^{***}$ \\ 
  & (0.017) & (0.019) & (0.017) & (0.019) \\ 
  & & & & \\ 
 N. Credits (z-score) & 0.029$^{***}$ & $-$0.007 & 0.031$^{***}$ & $-$0.004 \\ 
  & (0.011) & (0.012) & (0.011) & (0.012) \\ 
  & & & & \\ 
 Forum Writing (z-score) & 0.075$^{***}$ & 0.075$^{***}$ &  &  \\ 
  & (0.017) & (0.016) &  &  \\ 
  & & & & \\ 
 Forum Reading (z-score) & 0.020 & 0.033$^{*}$ &  &  \\ 
  & (0.021) & (0.020) &  &  \\ 
  & & & & \\ 
 Node Degree (z-score) & 0.133$^{***}$ & 0.084$^{**}$ & 0.175$^{***}$ & 0.157$^{***}$ \\ 
  & (0.024) & (0.038) & (0.021) & (0.037) \\ 
  & & & & \\ 
 Content Exposure [Low] &  & $-$0.004 &  & $-$0.016 \\ 
  &  & (0.037) &  & (0.037) \\ 
  & & & & \\ 
 Content Exposure [Medium] &  & 0.006 &  & $-$0.001 \\ 
  &  & (0.030) &  & (0.031) \\ 
  & & & & \\ 
 High School GPA (z-score) & 0.119$^{***}$ & 0.121$^{***}$ & 0.121$^{***}$ & 0.121$^{***}$ \\ 
  & (0.011) & (0.012) & (0.011) & (0.012) \\ 
  & & & & \\ 
 Age (z-score) & 0.005 & $-$0.002 & 0.004 & $-$0.003 \\ 
  & (0.008) & (0.009) & (0.008) & (0.009) \\ 
  & & & & \\ 
 Node Degree (z-score) x Content Exposure [Low] &  & 0.044 &  & 0.023 \\ 
  &  & (0.039) &  & (0.039) \\ 
  & & & & \\ 
 Node Degree (z-score) x Content Exposure [Medium] &  & $-$0.013 &  & $-$0.035 \\ 
  &  & (0.033) &  & (0.033) \\ 
  & & & & \\ 
 Node Degree (z-score) x High School GPA (z-score) & $-$0.024$^{**}$ & $-$0.023$^{*}$ & $-$0.025$^{**}$ & $-$0.023$^{*}$ \\ 
  & (0.011) & (0.012) & (0.011) & (0.012) \\ 
  & & & & \\ 
 Constant & 5.868$^{***}$ & 5.879$^{***}$ & 5.874$^{***}$ & 5.892$^{***}$ \\ 
  & (0.043) & (0.051) & (0.044) & (0.052) \\ 
  & & & & \\ 
\hline \\[-1.8ex] 
Random slope for node degree in each degree program & Yes & Yes & Yes & Yes \\ 
Degree program random intercept & Yes & Yes & Yes & Yes \\ 
High school random intercept & Yes & Yes & Yes & Yes \\ 
Commune random intercept & Yes & Yes & Yes & Yes \\ 
\hline
Observations & 2,296 & 1,697 & 2,296 & 1,697 \\ 
Log Likelihood & $-$990.039 & $-$644.503 & $-$1,005.517 & $-$658.443 \\ 
Akaike Inf. Crit. & 2,010.079 & 1,327.006 & 2,037.034 & 1,350.885 \\ 
Bayesian Inf. Crit. & 2,096.163 & 1,430.302 & 2,111.640 & 1,443.308 \\ 
\hline 
\hline \\[-1.8ex] 
\textit{Note:}  & \multicolumn{4}{r}{$^{*}$p$<$0.1; $^{**}$p$<$0.05; $^{***}$p$<$0.01} \\ 
\end{tabular} 
\end{table}

\begin{table}[!htbp] \centering 
  \caption{Hierarchical regression model adding content exposure for third-year students enrolled before 2018.} 
  \label{bottom_top3} 
\tiny 
\begin{tabular}{@{\extracolsep{5pt}}lcccc} 
\\[-1.8ex]\hline 
\hline \\[-1.8ex] 
 & \multicolumn{4}{c}{\textit{Dependent variable:}} \\ 
\cline{2-5} 
\\[-1.8ex] & \multicolumn{4}{c}{Average GPA 1-2020} \\  & Main model & Main model and content & Main model  & Main model and content \\ 
  & &  & (no forum controls) & (no forum controls) \\
\\[-1.8ex] & (1) & (2) & (3) & (4)\\ 
\hline \\[-1.8ex] 
 Sex [M] & $-$0.103$^{***}$ & $-$0.061$^{*}$ & $-$0.108$^{***}$ & $-$0.070$^{**}$ \\ 
  & (0.023) & (0.031) & (0.023) & (0.032) \\ 
  & & & & \\ 
 N. Credits (z-score) & 0.068$^{***}$ & 0.082$^{***}$ & 0.069$^{***}$ & 0.084$^{***}$ \\ 
  & (0.012) & (0.019) & (0.012) & (0.019) \\ 
  & & & & \\ 
 Forum Writing (z-score) & 0.036 & 0.037$^{*}$ &  &  \\ 
  & (0.023) & (0.022) &  &  \\ 
  & & & & \\ 
 Forum Reading (z-score) & 0.072$^{**}$ & 0.062$^{**}$ &  &  \\ 
  & (0.029) & (0.027) &  &  \\ 
  & & & & \\ 
 Node Degree (z-score) & $-$0.019 & 0.033 & 0.042 & 0.099$^{***}$ \\ 
  & (0.030) & (0.040) & (0.028) & (0.038) \\ 
  & & & & \\ 
 Content Exposure [Low] &  & $-$0.122$^{**}$ &  & $-$0.128$^{**}$ \\ 
  &  & (0.053) &  & (0.054) \\ 
  & & & & \\ 
 Content Exposure [Medium] &  & $-$0.060 &  & $-$0.067 \\ 
  &  & (0.045) &  & (0.046) \\ 
  & & & & \\ 
 High School GPA (z-score) & 0.081$^{***}$ & 0.080$^{***}$ & 0.083$^{***}$ & 0.081$^{***}$ \\ 
  & (0.015) & (0.020) & (0.015) & (0.020) \\ 
  & & & & \\ 
 Age (z-score) & $-$0.023$^{**}$ & $-$0.030$^{**}$ & $-$0.025$^{**}$ & $-$0.033$^{**}$ \\ 
  & (0.011) & (0.015) & (0.011) & (0.015) \\ 
  & & & & \\ 
 Node Degree (z-score) x Content Exposure [Low] &  & $-$0.005 &  & $-$0.017 \\ 
  &  & (0.048) &  & (0.049) \\ 
  & & & & \\ 
 Node Degree (z-score) x Content Exposure [Medium] &  & $-$0.032 &  & $-$0.050 \\ 
  &  & (0.038) &  & (0.039) \\ 
  & & & & \\ 
 Node Degree (z-score) x High School GPA (z-score) & 0.005 & 0.003 & 0.006 & 0.004 \\ 
  & (0.015) & (0.017) & (0.015) & (0.017) \\ 
  & & & & \\ 
 Constant & 5.578$^{***}$ & 5.575$^{***}$ & 5.582$^{***}$ & 5.587$^{***}$ \\ 
  & (0.045) & (0.061) & (0.045) & (0.061) \\ 
  & & & & \\ 
\hline \\[-1.8ex] 
Random slope for node degree in each degree program & Yes & Yes & Yes & Yes \\ 
Degree program random intercept & Yes & Yes & Yes & Yes \\ 
High school random intercept & Yes & Yes & Yes & Yes \\ 
Commune random intercept & Yes & Yes & Yes & Yes \\ 
\hline
Observations & 1,647 & 831 & 1,647 & 831 \\ 
Log Likelihood & $-$974.610 & $-$480.193 & $-$980.271 & $-$485.264 \\ 
Akaike Inf. Crit. & 1,979.220 & 998.387 & 1,986.543 & 1,004.527 \\ 
Bayesian Inf. Crit. & 2,060.321 & 1,088.117 & 2,056.830 & 1,084.812 \\ 
\hline 
\hline \\[-1.8ex] 
\textit{Note:}  & \multicolumn{4}{r}{$^{*}$p$<$0.1; $^{**}$p$<$0.05; $^{***}$p$<$0.01} \\ 
\end{tabular} 
\end{table}

\end{document}